\documentclass[aps,prl,twocolumn,groupedaddress,nofootinbib]{revtex4-1}
\usepackage{epsfig,amsmath,amssymb,graphics,color,calc}
\usepackage{mathtools}
\pdfoutput=1

\begin{document}
\title{Marginally Stable Equilibria in Critical Ecosystems}

\author{Giulio Biroli\textsuperscript{a,b,}, Guy Bunin\textsuperscript{c}, Chiara Cammarota\textsuperscript{d}}

\affiliation{\textsuperscript{a}Institut de physique th\'eorique, Universit\'e Paris Saclay, CEA, CNRS, F-91191 Gif-sur-Yvette, France, \textsuperscript{b}Laboratoire de Physique Statistique, \'Ecole Normale Sup\'erieure, CNRS, 
        PSL Research University, Sorbonne Universit\'es, 75005 Paris, France, \textsuperscript{c}Department of Physics, Technion-Israel Institute of Technology, Haifa 32000, Israel, \textsuperscript{d}Department of Mathematics, King's College London, London WC2R 2LS, UK}





\begin{abstract}
In this work we study the stability of the equilibria reached by ecosystems formed by a large number of species. The model we focus on are Lotka-Volterra equations with symmetric random interactions. Our theoretical analysis, confirmed by our numerical studies, shows that for strong and heterogeneous interactions 
the system displays multiple equilibria which are all marginally stable. 
This property allows us to obtain general identities between diversity and single species responses, which generalize and saturate May's bound. 
By connecting the model to systems studied in condensed matter physics, we show that the multiple equilibria regime is analogous to a critical spin-glass phase. This relation provides a new perspective as to why many systems in several different fields appear to be poised at the edge of stability and also suggests new experimental ways to probe marginal stability. 
\end{abstract}

\maketitle




Many complex systems in Nature organize in states that are poised just at the edge of stability. The growing evidence comes from physics \cite{wyart-muller-review}, biology \cite{bialek-critical}, ecology \cite{ellner1995chaos}, neuroscience \cite{neuro1,neuro2} and economy \cite{woodford-critical}.
One important common trait of all examples is that they are formed by strongly interacting units---species, neurons, agents and particles depending on the situation.  The possible explanations of such phenomenon are varied. 
They include the need for flexibility and adaptiveness to time-varying conditions \cite{bialek-critical,maritan-critical}, balance between functionality and stability \cite{maritan-critical}, self-organized criticality \cite{bak2013nature}, 
self-organized instability \cite{allesinasole}, and continuous constraints satisfaction \cite{wyart-muller-review}.\\
Here we address this problem focusing on generalized Lotka-Volterra (LV) equations. They provide a simple and general setting to study assemblies of interacting degrees of freedom; as such they are used in several fields \cite{bucci2014towards,faust2012microbial,may2007theoretical,goodwin1990chaotic}. In particular, they provide a canonical model for ecosystems, with growing connections to systems across biology \cite{bucci2014towards,faust2012microbial,eco-evo}. 
The study of stability of equilibria and their properties using LV equations and generalizations 
has become a very active research subject. Several important results were obtained recently; in particular
general techniques to count the number of equilibria and their properties 
have been developed \cite{fyod}, and criticality and glassiness have been found to be emergent properties 
of ecosystems \cite{kessler,kesslerlevine,mehta}. Our approach unifies these different perspectives and, by a mapping 
to condensed matter systems, reveal their generality beyond LV models.    
Henceforth, in order to describe it, we shall use the terminology employed in theoretical ecology.  

In the model we consider, an ecological community is assembled from a pool of available species.
We focus on the case relevant for the examples cited above, and in many other situations, when the number of species is large.
Since the detailed parameters of all interactions are not known in the majority of cases, and in any case not all details are expected to matter \cite{Barbier}, we follow the long tradition pioneered by May in ecology \cite{May} and Wigner in physics \cite{wigner1967random}, and sample the interactions randomly. However, 
we go beyond May's classical work since randomness is here introduced at the level of interactions between all possible species, while the community self-organizes by choosing which species are present. 
In other words, the number and identity of the species that are present in the community is selected {\em dynamically} \cite{macarthur,opper}. Understanding the emergent stability of the equilibria reached dynamically and its dependence on the external parameters is the main purpose of this work.  \\
We find, in agreement with \cite{guyPRE,kessler,mehta}, that when the interactions are weak or highly uniform, only one equilibrium is present and is determined mainly by self-regulation within each species. For stronger and more heterogeneous interactions, multiple equilibria emerge. Our main result is that when this happens,  all possible states of the system are close to be marginally stable for large number of species and this determines the diversity of the ecosystem, see Fig. 1. Marginal stability has several important consequences, in particular it leads to extreme susceptibility to small perturbations. This situation is referred as \textquotedblleft critical\textquotedblright\ in the physics literature \cite{chaikinlubensky}.
May famously suggested that complexity and interactions limit the stability of ecosystems \cite{May}. Our results provide a complementary perspective: complex ecological communities reduce dynamically their instability through a reduction of the possible number of surviving species, i.e. diversity, and eventually reach a marginally stable state saturating May's bound.
Since this phenomenon stems from a dynamical process, it holds for a broad range of system parameters. It is robust against a range of variations in the model, including different functional forms of responses and interactions, as well as noise. Although in many physical cases criticality emerges only at phase transitions, i.e. for very special values of the parameters, there also exist critical phases of matter which instead cover a wide portion of their phase diagram. By relating the LV model to systems studied in condensed matter physics, the multiple-equilibria regime is shown to be akin to a critical spin-glass phase. This suggests that the applicability of our results goes well beyond the LV model we consider and it offers a possible explanation of why so many different systems are found at the edge of stability: they are in a critical marginally stable phase. It also makes clear that this result, while general, is expected to have a well-defined regime of validity, as we shall explain at the end of this work. Finally it makes predictions on some distinctive features of the dynamical behaviour of ecological systems at criticality, which will be interesting to test.

\begin{figure}
\centering
\includegraphics[width=.8\linewidth]{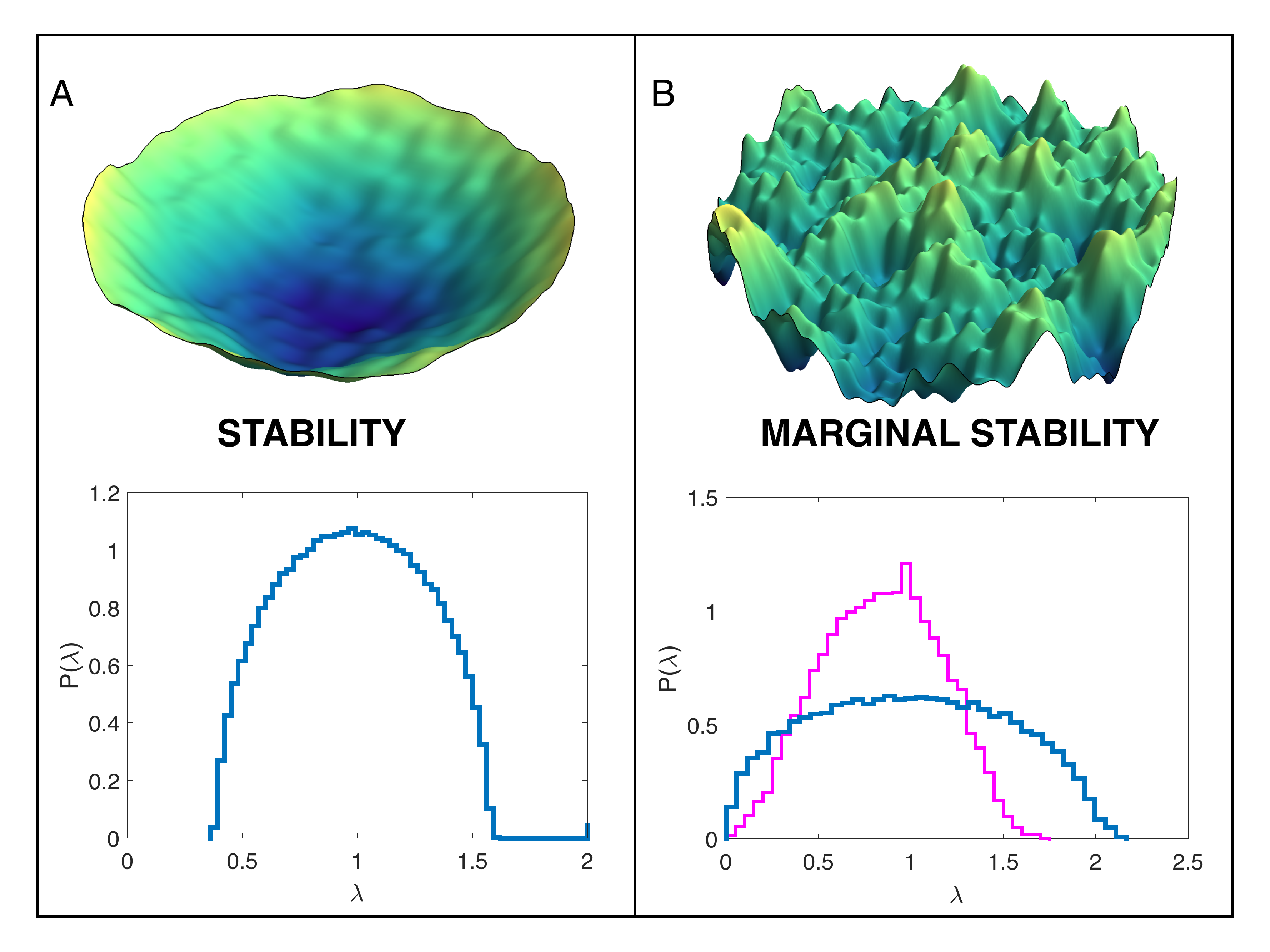}
\caption{Possible scenarios for the energy landscape associated to Lotka-Volterra dynamics. (A)\ There is only a single equilibrium, i.e. a unique global and local minimum, as illustrated by the cartoon of the energy landscape. The corresponding density $\rho(\lambda) $ of eigenvalues of the stability matrix associated with a given minimum (the Hessian) has a strictly positive support and the number of species in the community is strictly smaller than May's bound.  Here we show the numerical example obtained for the standard Lotka-Volterra model ($f(N)=1-N$, $r_i=K_i=1$ and $\mu=4, \sigma=0.5, S=400$). As explained in the text, for a large number of species, $\rho(\lambda)$ is in this case a shifted Wigner semi-circle.      
(B)\ The energy landscape is rugged: there are many equilibria and local minima, as illustrated 
by the cartoon of the energy landscape. The corresponding density $\rho(\lambda) $ of eigenvalues of the stability matrix associated with a minimum has a support whose left edge touches zero, corresponding to marginal stability, and the number of possible surviving species saturates May's bound, see Fig.3. 
Here we show the numerical example obtained for the standard Lotka-Volterra model ($f(N)=1-N$, $r_i=K_i=1$ and $\mu=4, \sigma=0.9, S=200$) in blue and for a different functional response ($f(N)=1-N-3/4(N-1)^2$, $r_i=K_i=1$ and $\mu=4, \sigma=0.5, S=200$) in magenta. In the former case $\rho(\lambda)$ is a shifted Wigner semi-circle, whereas in the latter it has a different shape.}
\label{fig:overview2}
\end{figure}

The Lotka-Volterra model we focus on is defined as follows. There are $S$ species in the regional pool,
whose abundance is $N_{i}\ge 0$. The dynamical equations read%
\begin{equation}\label{LVeqs}
\frac{dN_{i}}{dt}=\frac{r_i}{K_i} N_i (K_i-N_i)
-N_i\sum_{j,\left(  j\neq i\right)  }\alpha_{ij}N_{j}
+\sqrt{N_{i}}\eta_i(t)+\lambda
\end{equation}
where $r_i$ is the intrinsic growth rate of species $i$, and $K_i$ is the carrying capacity. 
It corresponds to the equilibrium abundance to which species $i$ would self-regulate in absence of interaction. 
For sake of clarity, in the following we focus on the case 
where $r_i$ and $K_i$ are constants (set equal to one by rescaling the other parameters).   
Later, we shall consider the effect of variability in $r_i$ and $K_i$, and also different functional responses by replacing $N_i(K_i-N_i)$ with more general forms, such as $N_if(N_i)$.
The interaction between species is encoded in the matrix 
$\alpha_{ij}$. We also add a small (infinitesimal) immigration rate $\lambda$ 
to ensure that all invadable species exist\footnote{A species dependent $\lambda_{i}$ 
would not change the results of our analysis.}. Finally, $\eta_i(t)$ is a white noise with variance  $2\omega^2$, and $\sqrt{N_{i}}$ captures
the scaling of demographic noise\footnote{We use Ito's convention for the multiplicative noise since it 
correctly captures the fact that a species with $N_i=0$ remains at zero abundance also in presence of noise.}. 
We consider a symmetric interaction matrix $\alpha_{ij}=\alpha_{ji}$, corresponding to competitive (or weakly mutualistic) interactions; we will discuss in the conclusion the effect of asymmetry.  
Except for this constraint, no additional structure (such as trophic levels or space) is included, and the entries $\alpha_{ij}$ are taken to be independent identically distributed random variables. Note that, as already anticipated above, the assumption on the randomness is done at the level of the pool and not of the community. 
The random variable $\alpha_{ij}$ can be drawn from any distribution without long tails, all that matters are its mean and variance. It turns out that the parameters that play a role in the final theory are the average number of links, $C$, per site and the first two moments of $\alpha_{ij}$ though the combination $\mu=C \rm{mean} [\alpha_{ij} ]$ and $\sigma^{2}=C \rm{var}[ \alpha_{ij}]$. For the sake of clarity, we now focus on the case $C=S$ in which all species interact, extensions are discussed at the end of this paper.\\
The LV eqs. can be rewritten in a way that makes their relationship with stochastic equations 
studied in physics more transparent: 
\begin{equation}\label{lang}
\frac{dN_{i}}{dt}=-N_i\left[ \nabla_{N_i}V_i(N_i)
+\sum_{j,\left(  j\neq i\right)  }\alpha_{ij}N_{j}\right] 
+\sqrt{N_{i}}\eta_i(t)+\lambda
\end{equation}
where the "potential" $V_i(N_i)$ 
is equal to $\frac{r_i}{K_i}(-K_i N_i+N_i^2/2)$. Without noise these eqs. 
admit a Lyapounov
function
\[
L=-\sum_i V_i(N_i)-\frac 1 2 \sum_{i\ne j}\alpha_{ij} N_iN_j+\lambda \sum_i\log N_i \,.
\]
In presence of noise, eqs. (\ref{lang}) are generalized Langevin equations. In the SI we show that they represent equilibrium dynamics of a thermal system with temperature $T=\omega^2$ and characterized by 
the following effective Hamiltonian, or energy, $H=-L+\sum_i T \log N_i$. As a consequence, the long-time stationary probability distribution is the Boltzmann law: $P=e^{-H/T}/Z$, where the partition function $Z$
guarantees the normalization. This result reveals that understanding the equilibria and the dynamics associated with the LV eqs. (\ref{LVeqs})  can be exactly reformulated as a problem of statistical physics of thermal disordered systems, in which the $N_i$ represent the degrees of freedom interacting via random couplings $\alpha_{ij}$ and subjected to individual potentials\footnote{We restrict our analysis 
to the case $T<\lambda$, otherwise all species go extinct at long times.} $V_i(N_i)+(T-\lambda) \log N_i$. 
Therefore,  we can deduce properties of the equilibria reached dynamically by a thermodynamical analysis. In particular, without noise, i.e. at zero temperature, the equilibria correspond to the minima of the energy function, i.e. to the ground state and the metastable states.
In order to obtain the full solution of this model we shall use tools developed in statistical physics of disordered systems, including replica computations, discussed in detail in the SI. 

The phase diagram without noise, i.e. at zero temperature, and for small migration, $\lambda\rightarrow 0^+$, has been obtained in \cite{guyPRE}.
We reproduce it in the SI for completeness (Fig. \ref{phasediag}) and show that it coincides, as it should, with the one obtained by the replica method. One finds that when interaction strengths are all identical, all species coexist in the community when $\mu>0$\footnote{The range $-1<\mu<0$ is not of interest for this work because the multiple equilibria phase we want to study is not present.}. By increasing the variability in the interactions, more and more species are driven out of the community. They are characterized by $N_i=0$ for $\lambda \rightarrow 0^+$ (we will call them ``extinct'' henceforth). The equilibrium reached dynamically is stable under perturbations, that is by changing $V_i(N_i)\rightarrow V_i(N_i)-\xi_i N_i$, and there is a gap in the spectrum of the corresponding stability matrix\footnote{This matrix, determining the stability under changes that affect the carrying capacities, is different from the one governing stability under the demographic noise $\eta_i$.}
 \cite{guyPRE}, see Fig. \ref{fig:overview2}(A). Note that in this regime the final community composition is unique, independent of the assembly history, e.g. , the initial conditions for the dynamics.
This picture persists up to $\sigma_c=1/\sqrt{2}$. By increasing randomness in the interactions above $\sigma_c$ a transition
to another regime, which is sharp for large systems, takes place.
\begin{figure}
\centering
\includegraphics[width=1\linewidth]{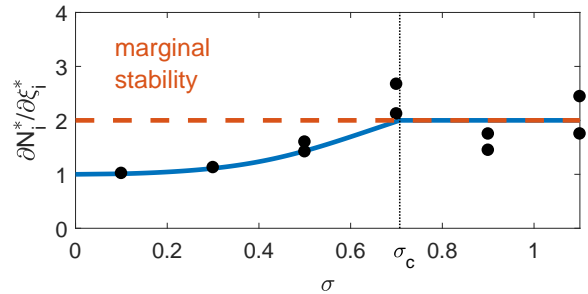}
\caption{Single species response as a function of $\sigma$ for two given species in an equilibrium reached dynamically for the LV model with $r_i=K_i=1$ and $S=400$. 
The numerical results follow the continuous line, which is the analytical prediction valid in the large $S$ limit
for all species. The fluctuations are finite $S$ effects. The single species response first increases with $\sigma$ and then sticks to the value $2$ in the whole marginally stable phase.}
\label{fig2}
\end{figure}
Our purpose is to study the phase reached when crossing the transition. Henceforth we continue to focus on the zero-noise case, the effect of the demographic noise is discussed at the end of the paper.\\
In physics terms, the single equilibrium regime corresponds to a ``paramagnetic phase'' where the zero-temperature values of the degrees of freedom $N_i$ are mainly fixed by the external potential 
$V_i(N_i)$ and a mean-field anti-ferromagnetic (competitive) interaction. By increasing the randomness in the $\alpha_{ij}s$ the system undergoes a zero temperature phase transition toward a spin-glass phase, characterized by many local minima of the energy (or maxima of the Lyapounov) function and, hence, multiple equilibria.
We have used the replica method to study it (see SI) and found that the regime with multiple equilibria corresponds, technically, to a full replica symmetry breaking solution. 
On the basis of all previous analysis of mean-field\footnote{Here the term mean-field refers to the fact that the underlying interaction network is fully connected, and not (as often used in ecology) that all interactions are identical in strength.} spin-glasses  \cite{spinglassbeyond,cukureview}, we can then make general statements about the regime with multiple equilibria. \\ 
First, it is characterised by a large number of equilibria. These equilibria are minima of the energy, separated by regions with higher energies that form what are called barriers. 
The lowest equilibria are typically separated by barriers that diverge in the large $S$ limit, while the higher ones by barriers of order of one, i.e. that do not scale with $S$ \cite{aspelmeier}. Second, and central to our discussion, all minima display a stability matrix characterized by arbitrary small eigenvalues for large $S$, i.e. minima are marginally stable and characterized by flat directions in the energy landscape at quadratic order. The ground state has this property and, naturally, the higher energy local minima too. \\
\begin{figure}
\centering
\includegraphics[width=1\linewidth]{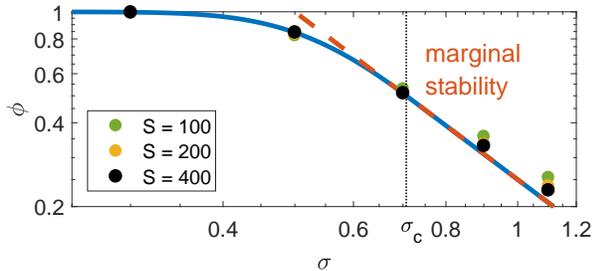}
\caption{Diversity $\phi$ in the standard LV case, $f(N)=1-N$ and $r_i=K_i=1$, as a function of $\sigma$ and for $S=100,200,400$. The diversity hits and sticks to the May bound throughout the entire multiple equilibria phase. 
The difference with the analytical predictions are finite $S$ effects.}
\label{fig3}
\end{figure}
We now explain the main findings of our thermodynamic analysis and relate them to 
random matrix theory results, see SI for details. 
The two main observables we focus on are: (i) $ \frac{\partial N_{i}^{\ast}}{\partial\xi_{i}^{\ast}}$, which is the response of a single species to a perturbation $V(N_i)\rightarrow V(N_i)-\xi_i N_i$, where the star indicates that only non-extinct species are considered and (ii) $\phi\equiv S^{\ast}/S$, which is the fraction of species present in the community, called diversity in what follows.
For identical interaction strengths, i.e. at $\sigma=0$, all species coexist ($\phi=1$) and $ \frac{\partial N_{i}^{\ast}}{\partial\xi_{i}^{\ast}}=1$. Increasing $\sigma$ we find that $ \frac{\partial N_{i}^{\ast}}{\partial\xi_{i}^{\ast}}$ is constant across species and increasing. Concomitantly, the diversity decreases. As found in \cite{guyPRE}, at $\sigma_c=1/\sqrt{2}$ the system undergoes a sharp transition from the single to the multiple equilibria regime. This corresponds physically to a phase transition to the spin-glass phase. In this phase we find that for all equilibria and all species $i$,
\begin{equation}\label{mayeq}
\frac{\partial N_{i}^{\ast}}{\partial\xi_{i}^{\ast}}=\sqrt{\frac{1}{\phi_c \sigma_c^2} }=2\,\,\,\,\,\,{\rm and}\,\,\,\,\,\, \phi \sigma^2=\phi_c\sigma^2_c=\frac 1 4 \ ,
\end{equation}
where $\phi_c$ is the value of the diversity at the transition.
These identities hold throughout the entire multiple equilibria phase and they are consequences of the 
criticality of the spin-glass phase. We also find that for $\sigma>\sigma_c$ the number of equilibria, i.e. of energy landscape minima is exponential\footnote{The number of minima scales as $e^{hS}$ where $h$ goes to zero at the transition from one to multiple equilibria.  One cannot do general statements on its order of magnitude, since it depends on the external parameters and on the model, in particular the functional response $f(N)$. When comparing to numerical and experimental results, it is important to keep in mind that it can be small, as we found for instance for the standard LV model. In consequence even for rather large $S\sim 100$ the number of equilibria may be modest, see SI.} in $S$. 
Figs. 2 and 3 confirm our analytical predictions by showing respectively numerical results for $\frac{\partial N_{i}^{\ast}}{\partial\xi_{i}^{\ast}}$ and $\phi$ corresponding to a given equilibrium reached dynamically.   \\
The second identity in (\ref{mayeq}) corresponds to a saturated form of May's original bound\footnote{The prefactor 4 comes from the symmetry of $\alpha_{ij}$ in the present model.}, $\mathrm{var}(\alpha_{ij})=1/(4S^*)=1/(4C^{\ast})$, where $C^*$ is the average number of interactions per surviving species \cite{May}. In order to reveal this connection with random matrix theory we focus on the $S^\ast \times S^\ast$ stability matrix $M^*$ associated to a given equilibrium, defined by the relation $(M^*)^{-1}_{ij}=\frac{\partial N_i^*}{\partial \xi_j^*}$. Using the fixed point equation corresponding to (\ref{LVeqs}) it is easy to check that 
\begin{equation}\label{SM}
M_{ij}^*=\delta_{ij}+\alpha_{ij}
\end{equation}
In this equation the indices $i,j$ have to be reduced to the surviving species since extinct species remain so if one adds an infinitesimal perturbation $\xi_j$, and do not contribute to the stability of the equilibrium reached dynamically\footnote{In the limit of small migration, $\lambda\rightarrow0^+$, extinct species are those that cannot invade: $\nabla_{N_i}V_i(0)+\sum_{j(\ne i)}\alpha_{ij} N_j^* > 0$. Adding an infinitesimal $\xi_i$ does not change this property and thus the species remains extinct.}. 
Following procedures developed for mean-field spin-glasses \cite{BM}, one can show (see SI) that the spectrum of $M_{ij}^*$ is identical for large $S$ to the one of a $S^\ast \times S^\ast$  matrix with independent identically distributed Gaussian off-diagonal entries having the same first and second moment of $\alpha_{ij}$.
This is by no means trivial since the equilibrium reached dynamically, and hence the identity of the surviving species, depend on $\alpha_{ij}$ and induce correlations in the off-diagonal 
elements of $M_{ij}^*$ \cite{guy}. The relation with random matrices implies that the eigenvalue density of the stability matrix is a Wigner semi-circle with 
support $[-2\sigma\sqrt\phi+1,2\sigma\sqrt\phi+1]$, as we indeed find numerically, see Fig. 1. Moreover, this directly connects (\ref{mayeq}), which holds in the entire spin-glass phase, to marginal stability.
We have therefore recovered May's original stability bound but in a saturated form: the number of surviving species, $S\phi$, is exactly the one guaranteeing that the system is poised at the edge of stability, similarly to what was proposed in the self-organized instability scenario \cite{allesinasole}.   

Let us now discuss extensions and the range of validity of our results. 
We have verified that our conclusions on the multiple equilibria regime continue to hold for several different convex functional responses $f(N)$, the standard $f(N)=N(1-N)$ being only an example among others, and with variability in the values of $r_i$ and $K_i$. 
This is a direct consequence of the properties of the spin-glass phase to which the multiple equilibria regime is related to. In these more general cases, the critical character of this phase is encoded in the following identity valid for the average of the square of the single species response in the whole multiple equilibria regime: 
\begin{equation}\label{gmayeq}
{\phi \sigma^2\left(
\frac{1}{S^*}\sum_{i=1}^{S^*}\left(  \frac{\partial N_{i}^{\ast}}{\partial\xi_{i}^{\ast}}\right)
^{2}\right)}=1\ .
\end{equation}
Note that eq. (\ref{gmayeq}) reduces to the first identity in eq. (\ref{mayeq}) when single species response are identical, as previously discussed.
We show in Fig. 4 the numerical results obtained for $f\left( N\right)=1-N-(N-1)^{2}/4$  confirming this prediction:  
the RHS of eq. (\ref{gmayeq}) is less than one in the single equilibrium phase, it reaches one at the transition and then remains stuck to this value in the whole multiple equilibria phase.
As before, the criticality of the spin-glass phase implies marginal stability.  
Indeed, similarly to the standard LV case, the spectrum of the stability matrix $M_{ij}^*=V''(N_i^*)\delta_{ij}+\alpha_{ij}$ 
can be shown to be identical for large $S$ to that of a $S^\ast\times S^\ast$ matrix with independent identically distributed Gaussian off-diagonal entries having the same first and second moment of $\alpha_{ij}$, and 
independent identically distributed diagonal entries with the same statistics of $V''(N_i^*)$. 
As we show in the SI, the condition that the left edge of the eigenvalues density touches zero for this class of random matrices is ${\phi \sigma^2\left(
\frac{1}{S^*}\sum_{i=1}^{S^*}\left((M^*)^{-1}_{ii} \right)^{2}\right)}=1$, which using $ \frac{\partial N_{i}^{\ast}}{\partial\xi_{i}^{\ast}}=(M^*)^{-1}_{ii}$ turns out to be identical to eq. (\ref{gmayeq}). Therefore we obtain that the multiple equilibria regime is indeed generically characterized by marginal stability and, by doing so, we derive a new generalized version of May's bound (eq. (\ref{gmayeq})). Remarkably, these properties hold despite the fact that for this general class of random matrices the density of eigenvalues is no longer a shifted semi-circle and the singularity at the left edge is not necessarily a square root\footnote{The singularity at the edge depends on the statistics of the diagonal components $M_{ii}^*$, i.e. of $V''(N^*_{i})$. It is a square root if the distribution of the random variable $V(N^*_{i})$ approaches the left edge of the support slower than linearly, in the other cases it may 
be a square root or instead inherit the singularity of $V''(N^*_{i})$ at the left edge \cite{mathpaper}.}, as shown numerically in Fig. 1 for the $f(N)=1-N-3/4(N-1)^2$ case  (which has $f'(N)>0$ for some values of $N$, corresponding to an Allee effect). 
Note that for general $f(N)$ the phase diagram is modified compared to the standard LV model. For instance for $f(N)=1-N-3/4(N-1)^2$  the single equilibrium phase is absent even for infinitesimally small interactions. In conclusion, our investigations show that the class of $f(N)$ leading to the marginally stable multiple-equilibria phase, i.e. the phase boundaries of the critical spin-glass phase in the physics terminology, is quite large. Determining its boundaries is an important and interesting task that we leave for future studies. Based on previous results on mean-field glassy systems \cite{cavagna}, it is possible that the property that the multiple equilibria  probed by the system are marginally stable is robust, even though the detailed properties of the landscape might be different depending on the shape of $f(N)$ \footnote{This is due to the fact that in many different situations, the most numerous minima, which are the ones with the biggest basins of attraction, are marginally stable \cite{cavagna}. Therefore it is not necessary that all minima are marginally stable to find that the ones reached {\it dynamically} are like this. }. Another extension of our work worthy of future analysis concerns the role of the interactions network. As long as the connectivity $C$ per species is large and the underlying structure rather homogeneous, e.g. no fat tails in the distribution of the local connectivity, the mean-field approach we developed is a very good approximation. In consequence, our results are expected to hold also in this more 
general setting where  $C\gg 1$ but $C$ is not necessarily equal to $S$. 
As a first interesting extension one could consider LV models on random regular graphs \cite{RRG}.\\
\begin{figure}
\centering
\includegraphics[width=1\linewidth]{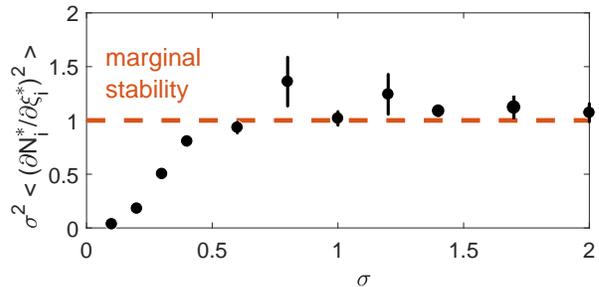}
\caption{Numerical test of the identity (\ref{gmayeq}) combining diversity $\phi$ and single species response valid throughout the multiple equilibria phase, for $f\left( N\right)=1-N-N^{2}/4$, $r_i=K_i=1$ and $S=400$. The fluctuations are finite $S$ effects.}
\label{fig4new}
\end{figure}
The properties (and the existence) of the multi-equilibria phase continue to hold also in presence of small noise. 
On the basis of previous studies on mean-field spin-glasses \cite{aspelmeier,cukureview}, we can state several general facts. In presence of small noise the system moves around between multiple dynamical states. These correspond to low temperature spin-glass states associated with the local minima discussed above. Only the low energy minima are able to trap the dynamics of the ecosystem over long periods of time, while the ones higher in energy are instead separated by small energy barriers; the transitions between them are so frequent that their identities as separate states disappear even for small noise. The stability of these dynamical states can described by the matrix ${\mathcal M}_{ij}$, which is a generalization of the stability matrix and is defined as the (matrix) inverse of $\frac{\partial \langle N_i \rangle}{\partial \xi_j}$ ($\langle \cdot \rangle$ denotes the average over the noise). As the stability matrix in absence of noise, ${\mathcal M}_{ij}$ is positive definite and has arbitrary small eigenvalues for large $S$, thus leading to marginal stability. 
Physically, this is a consequence of the fact that the spin-glass phase
is not destroyed by small thermal fluctuations and is critical over a finite range of temperatures \cite{spinglassbeyond}. \\
In summary, by mapping the LV models to thermal disordered systems and studying their thermodynamics, we find that marginal stability is a property of all communities that are reached dynamically by an ecosystem in the multiple equilibria phase.  Eq. (\ref{gmayeq}), combined with our random matrix analysis, relates this property to the single species response. This is the main result of our work: it represents an exact statement of May's stability bound \cite{May}, with three notable differences: (1) it follows from an exact analysis of the communities reached dynamically rather than from a-posteriori assumptions on the stability equations, it is thus a property of the emergent community; (2) it is saturated, with an equality rather than an inequality; (3) it is more general, allowing to incorporate non-linear $f\left( N\right)  $.

Having established that marginal stability is a generic property of the multiple equilibria phase, we now discuss some of its consequences and propose measurable tests. 
The most striking effects are expected to appear in dynamical phenomena. 
Again, previous results on dynamics of mean-field spin glasses provide useful guidelines \cite{cukureview}. 
In particular, starting the LV dynamics from random initial conditions one expects slow relaxations toward the minima and history dependence for large $S$. Both phenomena are tightly linked to marginal stability which results in flat directions in configurations space. The response to perturbations is also expected to be very unusual: marginal stability should lead to strong and wildly 
fluctuating non-linear responses \cite{biroliurbani} and avalanches of extinctions and invasions \cite{wyart-muller-review}
\footnote{These are different from cascades in trophic systems (e.g., \cite{pace1999trophic}), 
as here no trophic structure is included.}.
Working out the relevance of this phenomena in various ecological contexts certainly merits future research.
The results we found for symmetric interactions have also important consequences for cases where $\alpha_{ij}$ are asymmetric. Indeed, given that the multiple equilibria are marginally stable, we expect that adding asymmetry leads immediately to a chaotic behavior in which the system moves among the different regions of configurations space corresponding to the vestige of those equilibria \cite{berthierkurchan,cugliandolokurchanledoussal,sompolinsky}.\\
\begin{figure}
\centering
\includegraphics[width=1\linewidth]{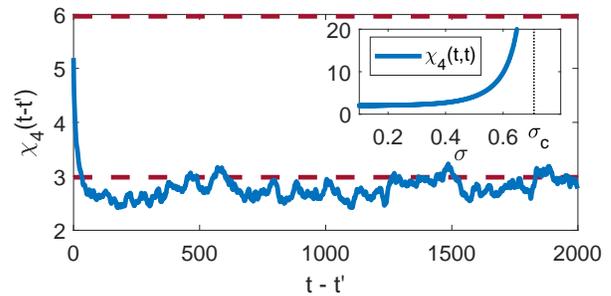}
\caption{We show $\chi_4(t,t')$ in the single equilibrium phase for the standard LV model ($f(N)=1-N$, $r_i=K_i=1$), $\sigma=0.55$, $S=400$ and $T=10^{-5}$.  
The top and bottom dashed lines show the values of $2\chi_{SG}$ and $\chi_{SG}$ respectively. Inset: Analytical prediction for $\chi_4(t,t)=2\chi_{SG}$ as a function of $\sigma$ approaching the transition toward the multiple equilibria phase. $\chi_{SG}$ diverges for $\sigma \rightarrow \sigma_c$.}
\label{figchi4}
\end{figure}
Throughout this paper, we stressed that the unusual properties of the multiple equilibria phase are related to the criticality of the corresponding spin-glass phase. In the following, we show that this relationship also suggests new ways to test for marginal stability. Criticality corresponds to a state in which the microscopic degrees of freedom are all strongly correlated, 
which naturally leads to singular responses. The properties of the stability matrix in the multiple equilibria phase are one facet of this phenomenon; diverging fluctuations are another. Note, however, that simple fluctuations such as 
$\overline{\langle N_i^2\rangle-\langle N_i\rangle^2}$ or its time-dependent counterpart $C(t,t')=
\overline{\langle N_i(t)N_i(t')\rangle-\langle N_i(t)\rangle\langle N_i(t')\rangle}$ do not capture criticality (the overbar denotes the average across the species). 
One has to focus on other kinds of correlations. See SI for detailed computations on this point and the following.
As a matter of fact, in the LV model we considered, which can be mapped onto thermal dynamics, diverging responses and fluctuations are exactly related by the fluctuation-dissipation relation $\frac{\partial \langle N_i \rangle}{\partial \xi_j}=\omega^{-2}(\langle N_i N_j \rangle-\langle N_i\rangle \langle N_j \rangle)$. On this basis, and following previous work 
on glassy systems \cite{dynhetbook}, we propose to probe criticality, or the collective nature of the equilibria and more generally of the dynamical states by looking at a fourth-order correlation function, called $\chi_4(t,t')$,  which reads: 
\[
\chi_4(t,t')=\frac 1 S\sum_{ij}\left[\langle \delta N_i(t) \delta N_i (t') \delta N_j(t) \delta N_j(t') \rangle-\right.
\]
\[\left.\qquad \,\,\,\,-
\langle \delta N_i(t) \delta N_i (t') \rangle \langle \delta N_j(t) \delta N_j(t') \rangle\right]
\] 
where $\delta N_i(t)=(N_i (t) -\langle N_i(t)\rangle)/\sqrt{C(t,t)}$. 
This function allows one to measure to what extent species are dynamically correlated and is therefore 
a way to quantitatively test criticality and marginal stability. In the LV model and for $\sigma<\sigma_c$ 
the dynamics becomes stationary after a short transient. In this case $\chi_4(t,t')$ depends on $t-t'$ only. 
Its long time limit ($t-t'\gg1$) is equal to:
\[
\chi_{SG} =\frac{1}{S{\overline{\langle N_i^2 \rangle_c}^2}}\sum_{i,j} [\langle N_i N_j \rangle-\langle N_i\rangle \langle N_j \rangle]^2,
\]
whereas for $t=t'$ and small temperatures $\chi_4(t,t)$ is equal to twice $\chi_{SG}$. 
As shown in Fig. 5, where we check these analytical predictions by numerical simulations for $S=400$,
$\chi_4(t-t')$ is a decreasing\footnote{The shape of $\chi_4(t,t')$
is different from the one found in glassy systems. The main difference which lies in the different nature of the degrees of freedom is that $\chi_4(t,t)$ becomes large approaching the transition, whereas in glassy systems it is featureless.} function of $t-t'$. 
In the condensed matter theory context $\chi_{SG}$ is known as the spin-glass susceptibility and is known to diverge in the entire spin-glass phase. We indeed recover this result and connect it to marginal stability since 
\[
\chi_{SG} = \left(\frac{\partial \langle N_i \rangle }{ \partial \xi_i }\right)^{-2}\int d\lambda \frac{\rho(\lambda)}{\lambda^2}\sim S^{1/3}
\]
where $\rho(\lambda)$ is the density of eigenvalues of the matrix ${\mathcal M}_{ij}$. 
The divergence of $\chi_{SG}$ as $S^{1/3}$ comes from the fact that the minimum eigenvalue of ${\mathcal M}$ scales as $S^{-2/3}$ \cite{majumdar}. In the inset of Fig. 5 we show the behavior of $\chi_{SG}$ obtained analytically in the large $S$ limit. 
Note that if the singularity of $\rho(\lambda)$ is milder than a square root, as it is the case for example for $f(N)=1-N-3/4(N-1)^2$, then one needs to consider high-order moments. The bottom line of this discussion is that if data on the time-dependence of abundances is available, the function $\chi_4(t,t')$ allows one to measure to what extent species dynamics are correlated and test directly for criticality and marginal stability. 
Even though in the simple LV case, measuring $\chi_{SG}$ would be sufficient for that purpose, measuring the time dependent four-point function $\chi_4(t,t')$ is the way to go in order to obtain information in more general cases, which may be neither stationary, nor related to thermal equilibrium dynamics.
    

In conclusion, our analysis of Lotka-Volterra equations in the limit of large species offers an explanation of why many systems in Nature are poised at the edge of stability: we have shown that when 
the parameters of an ecosystem cross the limit of stability, the system dynamically self-adapts to remain exactly marginally stable. It does so reducing the number of species in such a way to saturate May's bound, which therefore emerges as a result of a dynamical process. 
This leads to a whole critical phase with multiple marginally stable equilibria, which is expected to be present 
for several different models and to display highly non trivial dynamical behaviors. Its consequences can be relevant and important in many fields \cite{maritan-critical,allesina2011stability,ellner1995chaos,woodford-critical,JP-LV}.\\\\
\begin{acknowledgments}
We thank J.-F. Arnoldi, M. Barbier, J.-P. Bouchaud,  D.S. Fisher, B. Haegeman, J. Kurchan, T. Mora, V. Ros, E. Vanden-Eijnden, P. Vivo, A. Walczak for useful discussions. 
This work was partially supported by the grant from the Simons Foundation ($\sharp$454935, Giulio Biroli). 
\end{acknowledgments}

\bibliography{pnas-sample}

\clearpage
\onecolumngrid
\appendix
\section{Supporting Information (SI)}

\subsection{Numerics}
Simulations of Eq. 1 in the main text for Figs. 1-4 were done without noise ($T=0$). This results in a set of coupled ODEs, that where run with $\lambda =10^{-18}$ and terminated when $d\left( \ln N_{i}\right) /dt<10^{-10}$ for all $i$. To calculate the spectra in Fig. 1 and the diversity in Fig. 3, species were considered to have positive $N_{i}$ if $N_{i}>10^{-14}$ at the final time. In Figs. 1-4, each data point used 3000 runs, where for each one a new matrix $\alpha $ was sampled. \\
The noisy simulation in Fig. 5 was run with a simple Ito discretization of the stochastic differential equation: $\Delta N_{i}=A_{i}\Delta t+\sqrt{B_{i}\Delta t}$, where $A_{i}$ are the noiseless terms of $dN_{i}/dt$ and $B_{i}=2TN_{i}$. The simulation was run with $T=10^{-5}$.
\subsection{Spin glasses}
Many of the theoretical methods used in this work were originally developed to study disordered systems in physics, and in particular spin-glasses. Here we make some brief comments on such systems and how they relate to the present problem.\\
The behavior described in this work requires {\it variability} in the interaction strengths $\alpha_{ij}$ (as measured by their standard deviation $\sigma$). In physics, systems where interactions between the constituents exhibit analogous variations are known as ``disordered systems''. In particular, a {\it spin glasses} is systems where magnetic interactions vary between the different pairs of atoms. Models of such systems, starting with \cite{edwards1975theory}, traditionally use binary variables to model the state of each magnetic spin $\sigma\in\{-1,1\}$, while the complex interactions were modeled as i.i.d. Gaussian random variables.
For example, the first spin glass model \cite{edwards1975theory} is the Edwards and Anderson model where the Hamiltonian (the energy of a state) is given by
$$H_{\rm SK}=-\sum_{(i,j)}J_{i,j}\sigma_i\sigma_j \ ,$$
where only terms involving pairs of nearest neighbours $(i,j)$ are included in the sum, and $J_{i,j}$ are randomly sampled with zero mean. 
The first model to be solved is its fully connected analog, the Sherrington Kirkpatrick (SK) model where all pairs of spins are taken to interact \cite{sherrington1975solvable,kirkpatrick1978infinite}. 
A number of other spin glass models were considered along the years \cite{spinglassbeyond,castellani2005spin,elderfield1983spin,monasson1996entropy}. Among the most interesting ones, we mention those where interactions involve $p>2$ variables at a time, called $p$-spin models \cite{castellani2005spin}
$$H_{p}=-\sum_{(i_1,i_2,\dots,i_p)}J_{i_1,i_2,\dots,i_p}\sigma_{i_1}\sigma_{i_2}\dots\sigma_{i_2} \ .$$
It turned out that spin glass models show interesting new phases and phase transitions and can be classified in different universality classes which correspond to different macroscopic behaviours.\\
The techniques developed for the solution of spin glasses are in general useful to describe systems with high level of {\it frustration} \cite{toulouse1987theory} (i.e. absence of optimal solutions for certain instances of the couplings). For this reason they are widely applied nowadays in different fields including condensed matter (magnetic systems, supercooled liquids), biology (protein folding, neural networks), social sciences, economics, computer sciences (optimization theory, machine learning). \\
The major obstacle that had to be tackled while dealing with the solution of spin glass models is represented by the task of performing disorder averages of quantities of physical interest to extract information on the macroscopic behaviour of the system.
The information about equilibrium is contained in the so-called free energy of the system
$$F=-\frac{1}{\beta}\overline{\log\sum_{\{\sigma_i\}}\exp[-\beta H]}^J \ ,$$
where the overline represents the average over difference instances of the disordered couplings, the inner sum runs over all the possible configurations of spins, and the argument of the logarithm is commonly called partition function $$Z=\sum_{\{\sigma_i\}}\exp[-\beta H] \ .$$\\
The operation of taking the average usually is reduced to perform a Gaussian integration.
This would have required little computation effort indeed had the logarithm not been on the way. 
Yet its presence cannot be neglected nor the operation of taking the average simply performed on the logarithm's argument (the last procedures is called {\it annealed} calculation but does not lead to the correct solution in the interesting regimes).
To keep the order of the operations and yet end up with an analytically tractable problem the so-called {\it replica trick} was introduced \cite{spinglassbeyond}.
It amounts to use the fact that 
$$\overline{\log x}=\lim_{n\rightarrow0}\frac{\log\overline{x^n}}{n} \ ,$$
which can be easily verified. 
Hence the free energy can be written in terms of the so called replicated partition function $Z^n$ as 
$$-\beta F=\lim_{n\rightarrow 0} \frac{\log \overline{Z^n}}{n}=\lim_{n\rightarrow 0} \frac{\log \overline{\{\sum_{\{\sigma_i\}}\exp[-\beta H]\}^n}}{n}$$
where the power $n$ can be interpreted, before the limit $n\rightarrow 0$ is taken, as we were focusing on $n$ independent copies of the same system in presence of a unique sample of random couplings. Averages over different realizations of the disorder are in this form straightforward.
The copies in the spin glass jargon are usually called {\it replicas}.\\
To give an intuition of the physical meaning of the results that can be obtained within replica computations we must remember that frustrated systems are usually characterized by a multi minima structure of the energy, or any equivalent cost function that might be of interest. This arrangement of minima is uniquely associated to any instance of the random couplings. The role of replicas is the one of revealing the main features of this multi minima structure by independently probing different minima. In fact one of the most important piece of information that comes out from a replica computation is the average width of equilibrium minima and the average distance between pairs of them, or more in general the hierarchical arrangements of minima in the space of configurations.
All this is contained in the structure of {\it overlap} $Q_{a,b}$ (or similarity, which accounts for the inverse of distance in the space of configurations) between pairs of replicas $a,b$:
\begin{eqnarray}
Q_{a,b}&=&\frac{1}{N}\sum_i^N\overline{\frac{\sum\limits_{\{\sigma_i^c\}}\exp\left[-\beta\sum\limits_c^nH_c\right]\sigma_i^a\sigma_i^b}{\sum\limits_{\{\sigma_i^c\}}\exp\left[-\beta\sum\limits_c^nH_c\right]}}^J \nonumber \\ \nonumber
&=&\frac{1}{N}\sum_i^N\overline{\langle\sigma_i^a\sigma_i^b\rangle_{\rm AR}}^J \ ,
\end{eqnarray}
where $\langle \rangle_{AR}$ denotes the measure over all replicas with the Boltzmann weight and after averaging over disorder (RHS of the first equation above). 
Along a typical replica computation a new conceptual obstacle arises when the free energy is rewritten in terms of an integral over all the possible choices of the overlap matrix. In the thermodynamic limit ($N\rightarrow \infty$) the Laplace method (or saddle point method) can be applied to evaluate the integral but it requires to find the overlap matrix that maximizes the argument of the integral. 
This operation requires the introduction of a good {\it ansatz} for the overlap matrix. The currently used scheme was proposed by Parisi \cite{parisi1979toward,parisi1980sequence} and subsequently proved to be the one providing the correct saddle point result \cite{talagrand2003generalized}. It is called Replica Symmetry Breaking (RSB) scheme and will be discussed in more details in the following sections. \\
Depending on the number of steps of breaking of the replica symmetry required to get a meaningful solution \cite{de1978stability} (i.e. stable in the replica space), we could end up working with Replica Symmetric (RS) scheme, one step Replica Symmetry Breaking ($1$RSB), $\infty$ steps Replica Symmetry Breaking (FRSB), just to mention the most relevant ones.
This differentiation allows to classify spin glass models and characterize the features of their relevant phases and phase transition.\\
It turns out that the ecological model we consider in this work, at large values of $\sigma$, is characterized by a FRSB solution. The FRSB solution represents, as stressed in the main text, a critical phase. 
From the technical point of view, it is marginally stable, meaning that within the Laplace method it corresponds to an extremum with vanishing small eigenvalues. \\

\subsection{Mapping to a thermal disordered system}
We assume the starting equation to be 
\begin{equation}
\frac{dN_i}{dt}=\frac{r_i}{K_i}N_i\left[f_i(N_i)+\xi_i-\sum_{j(\neq i)}\alpha_{ij}N_j\right]+\sqrt{N_i}\eta_i+\lambda_i
\end{equation}
with $\alpha_{i,j}$ a symmetric matrix, and $\eta_i$ a white noise: $\langle\eta_i(t)\eta_j(t')\rangle=2T\delta_{ij}\delta(t-t')$.
Lotka-Volterra (LV) classical system of equations for interacting species can be obtained by setting $f_i(N_i)=K_i-N_i$. \\
By scaling the variables $N_i$ by $\langle K_i\rangle=\mu_K$: $\widetilde{N_i}=N_i/\mu_k$ and consequently also the relevant parameters $\widetilde{K_i}=K_i/\mu_K$, $\widetilde{\lambda_i}=\lambda_i/\mu_K$, $\widetilde{T}=T/\mu_K$, $\widetilde{\xi}_i=\xi_i/\mu_K$, and also the function $\widetilde{f}_i(\widetilde{N}_i)=f_i(N_i)/\mu_K$ (which is true in the LV case) we get an identical dynamical equation in terms of the new variables and the old $\alpha_{ij}$, and $r_i$.
Without loss of generality then we set $\mu_K=1$ and forget about all the tildas.\\
We also prefer to define an interaction matrix $\theta_{ij}=r_i\alpha_{ij}/K_i=\rho\alpha_{ij}$. In doing this we assume that the ratio $\rho=r_i/K_i$ is $i$-independent so that $\alpha_{ij}$ and $\theta_{ij}$ are symmetric at the same time.\\
Finally we define the function $V(N_i)$ such that $-\nabla_{N_i}V_i(N_i)=\rho f_i(N_i)$ (in the LV case $V_i(N_i)=-\rho(K_iN_i-N_i^2/2)$), we set $\xi_i=0$ and $\lambda_i=\lambda$.\\
The dynamical equations then read 
\begin{equation}
\frac{dN_i}{dt}=-N_i\nabla_{N_i}V_i(N_i)-N_i\sum_{j(\neq i)}\theta_{ij}N_j+\sqrt{N_i}\eta_i+\lambda \ .
\label{DYNeq}
\end{equation}
We want to show that this equation admits an invariant probability distribution in terms of a Hamiltonian $H$. 
To do so we derive the corresponding Fokker-Planck equation \cite{gardiner1985stochastic}.
We consider a generic observable $O(\{N_j\})$ and the time derivative of its average over the thermal noise $\frac{d}{dt}\langle O(\{N_j\})\rangle$.
This derivative will obtain us the time derivative of the distribution of our variables originated by the thermal noise itself:
\begin{eqnarray}
\frac{d}{dt}\langle O(\{N_j\})\rangle&=&\frac{d}{dt}\int \prod_idN_i P(\{N_j\},t)O(\{N_j\}) \label{F-Peq} \\
&=&\int \prod_idN_i \frac{\partial}{\partial t}P(\{N_j\},t)O(\{N_j\}) \ . \nonumber
\end{eqnarray}
Adopting the Ito convention \cite{gardiner1985stochastic}, 
the LHS of the equation (\ref{F-Peq}) corresponds to
\begin{equation}
\frac{d}{dt}\langle O(\{N_j\})\rangle=\langle\sum_i\frac{\partial O}{\partial N_i}\frac{dN_i}{dt}\rangle + \frac{1}{2}\langle\sum_{i,j}\frac{\partial^2 O}{\partial N_i\partial N_j} \eta_i \eta_j\sqrt{N_iN_j}\rangle \ .
\end{equation}
In the Ito prescription variables are not correlated with noise at the same time so, also using equation (\ref{DYNeq}), the previous equation becomes
\begin{equation}
\frac{d}{dt}\langle O(\{N_j\})\rangle =
\langle\sum_i\frac{\partial O}{\partial N_i}\mathcal{D}(\{N_j\})
\rangle 
+T\langle\sum_{i}\frac{\partial^2 O}{\partial N_i^2} N_i\rangle \ ,
\end{equation}
where
$$
\mathcal{D}(\{N_j\})=-N_i\nabla_{N_i}V_i(N_i)-N_i\sum_{j(\neq i)}\theta_{ij}N_j+\lambda \ .
$$
Re-writing now the average over the noise as an average over $P(\{N_j\},t)$
\begin{eqnarray}
\dfrac{d}{dt}\langle O(\{N_j\})\rangle
& =& \int \prod_idN_i P(\{N_j\},t) \times \\
& {\displaystyle\sum\limits_i} &\left(\frac{\partial O}{\partial N_i}
\mathcal{D}(\{N_j\})
+T\frac{\partial^2 O}{\partial N_i^2} N_i \hspace{-0.1cm}\right) , \nonumber
\end{eqnarray}
and integrating by parts we get
\begin{eqnarray}
\hspace{-2.5cm}
& \dfrac{d}{dt} &\langle O(\{N_j\})\rangle=\int \prod_idN_i O(\{N_j\}) \times\\ \nonumber
& {\displaystyle\sum\limits_i} &\left\{ T\frac{\partial^2 }{\partial N_i^2} [P(\{N_j\},t)N_i] 
-\frac{\partial }{\partial N_i}\hspace{-0.1cm}\left[
\mathcal{D}(\{N_j\})
P(\{N_j\},t) \right] \hspace{-0.1cm}\right\} \ .
\end{eqnarray}
By comparison with equation (\ref{F-Peq}) we can now write the Fokker-Planck equation as
\begin{eqnarray}
& {\displaystyle \frac{\partial}{\partial t}} &P(\{N_j\},t)= \\
& {\displaystyle\sum\limits_i} &\left\{ T\frac{\partial^2 }{\partial N_i^2} [P(\{N_j\},t)N_i] 
-\frac{\partial }{\partial N_i}\hspace{-0.1cm}\left[\mathcal{D}(\{N_j\})P(\{N_j\},t) \right] \right\} \ . \nonumber
\end{eqnarray}
The equilibrium distribution must satisfy then 
\begin{equation}
0=\sum_i\left\{T\frac{\partial^2 }{\partial N_i^2} [P(\{N_j\},t)N_i] 
-\frac{\partial }{\partial N_i}\hspace{-0.1cm}\left[\mathcal{D}(\{N_j\})P(\{N_j\},t) \right] \right\} \ , \nonumber
\end{equation}
which is obtained by imposing $\forall i$
\begin{equation}
\frac{\partial P(\{N_j\},t)}{\partial N_i} =
\left[\frac{\mathcal{D}(\{N_j\})}{T}-1\right]\frac{P(\{N_j\},t)}{N_i} \ .
\end{equation}
By asking that this equilibrium distribution is also of the form
\begin{equation}
P=Z^{-1}\exp(-\beta H) \ ,
\end{equation}
with $Z$ the usual normalizing partition function,
we obtain
\begin{equation}
\frac{\partial P}{\partial N_i}=-P\beta \frac{\partial H}{\partial N_i}
\end{equation}
and hence that 
\begin{equation}
\frac{\partial H}{\partial N_i}=-\frac{1}{N_i}[\mathcal{D}(\{N_j\})-T]
\end{equation}
so finally 
\begin{equation}\label{Ham}
H=\sum_i V_i(N_i)+\sum_{(i,j)}\theta_{ij}N_iN_j+(T-\lambda)\sum_i\ln(N_i)  \ .
\end{equation}
Note that having $\lambda>T$ (even when $T\rightarrow0$) is a fundamental element to have a regularized  $P(\{N_j\})$ at small $N_j$. \\
In conclusion the original dynamical equations describe the dynamical evolution of a system whose thermodynamics is determined by the Hamiltonian just obtained.
\subsection{Replica computation}
We use a replica approach to analyze the thermodynamics of the system characterized by the Hamiltonian (\ref{Ham}).
Recall that $\theta_{ij}$ are assumed to be Gaussian distributed with mean $\rho\mu/S$ and variance $\rho^2\sigma^2/S$.
We evaluate the free-energy of the system by applying the replica trick
to perform sample averages \begin{equation}
-\beta F=\lim_{n\rightarrow 0} \frac{\ln \overline{Z^n}}{n} \ .
\end{equation}
We then evaluate the replicated partition function as
\begin{equation}
\overline{Z^n}=\overline{\int \hspace{-0.2cm}\prod_{i,(ij)}\hspace{-0.1cm}d N_i^a d\theta_{ij} \exp\hspace{-0.1cm}\left(-\sum_{(ij)}\frac{(\theta_{ij}-\rho\mu/S)^2}{2\rho^2\sigma^2/S}-\beta H(\{N_i\})\hspace{-0.1cm}\right)}^V ,
\end{equation}
where the average over the disorder contained in $V$ remains to be done, and
which, through standard replica manipulations \cite{spinglassbeyond}, becomes
\begin{equation}
-\beta n F=\ln \overline{\int \hspace{-0.2cm}\prod_{a,a<b}\hspace{-0.1cm}dQ_{ab}dQ_{aa}dH_{a}
\exp\left[S \mathcal{A}(Q_{a,b},Q_{a,a},H_a)
\right]}^V 
\end{equation}
with 
\begin{eqnarray}
\mathcal{A}(Q_{a,b},Q_{a,a},H_a)& = &-\rho^2\sigma^2\beta^2\sum_{a<b}\frac{Q_{ab}^2}{2}-\rho^2\sigma^2\beta^2\sum_a\frac{Q_{aa}^2}{4} \nonumber \\ 
& + &\rho \mu \beta\sum_a   \frac{H_a^2}{2}+\frac 1 S \sum_i \ln Z_i 
\end{eqnarray}
In the last equation the effective on-site partition function $Z_i=\int \prod_{a}dN_i^a \exp\left(-\beta H_{eff}(\{N^a\}_i)\right)$ is obtained 
from an effective Hamiltonian
\begin{eqnarray}
H_{eff}(\{N^a\}_i)& = &-\beta \rho^2 \sigma^2 \sum_{a<b}N_i^a N_i^b Q_{ab}-
\beta \rho^2 \sigma^2 \sum_{a}(N_i^a)^2 \frac{Q_{aa}}{2} \nonumber \\
& + & \sum_{a} \rho \mu N_i^aH_a
+ V_i(N_i^a)+(T-\lambda)\log N_i^a
\end{eqnarray}
where the order parameters satisfy the self-consistent equations
\[
Q_{ab}=\frac 1 S \sum_i \overline{\langle N_i^a N_i^b \rangle_{\rm AR}}^V\,,\ \ \ \ \  
H_{a}=\frac 1 S \sum_i \overline{\langle N_i^a  \rangle_{\rm AR}}^V
\]
and the averages over configurations of all replicas (AR) are performed with the effective Hamiltonian $H_{eff}$.
\subsection{Replica Symmetric Solution}
Replica symmetry is correct when only a single equilibrium (or minimum) in the (free-)energy landscape governs the thermodynamic behavior. 
Formally, one assumes: 
\[
Q_{ab}=q_0 \,\forall a \ne b\qquad ,\qquad Q_{aa}=q_D  \,\forall a\qquad ,\qquad H_a=h \,\forall a \ .
\] 
The free-energy expression becomes in this case 
\begin{equation}
-\beta n F=\ln \overline{\int dq_0dq_Ddh
\exp\left[S\mathcal{A}(q_0,q_1,h)
\right] }^V
\label{FERS}
\end{equation}
\begin{eqnarray}
\mathcal{A}(q_0,q_1,h)& = &-\rho^2\sigma^2\beta^2\frac{n(n-1)}{4}q_0^2
-\rho^2\sigma^2\beta^2\frac{n}{4}q_D^2 \\
& + &\rho \mu \beta\frac{n}{2}h^2+\frac 1 S \sum_i \ln Z_i
\end{eqnarray}
with an effective Hamiltonian for $Z_i$
\begin{eqnarray}
H_{eff}(\{N^a\}_i)& = &-\frac{\beta \rho^2 \sigma^2}{2} q_0\hspace{-0.1cm}\left(\hspace{-0.1cm}\sum_{a}N_i^a\hspace{-0.1cm}\right)^2\hspace{-0.2cm} 
-\frac{\beta \rho^2 \sigma^2}{2} (q_D-q_0)\sum_{a}{N_i^a}^2 \nonumber \\
& + &  \sum_{a}\rho \mu hN_i^a
+ V_i(N_i^a)+(T-\lambda)\log N_i^a \ .
\end{eqnarray}
To decouple replicas we exploit standard properties of Gaussian integrals and we get 
\[
Z_i=\hspace{-0.1cm}\int\hspace{-0.1cm}\frac{dz_i}{\sqrt{2\pi}} \exp\hspace{-0.1cm}\left[-\frac{z_i^2}{2}\right]\int \prod_a dN_i^a\exp\left[-\beta\sum_aH_{\rm RS}(N_i^a,z_i)\right]
\]
with
\begin{eqnarray}
H_{\rm RS}(N_i,z_i)& = &-\rho^2 \sigma^2 \beta(q_D-q_0) \frac{N_i^2}{2}+\left(\rho \mu h-z_i \rho \sqrt q_0 \sigma\right)N_i \nonumber \\
& + &V_i(N_i)+(T-\lambda)\log N_i \ .
\end{eqnarray}
By maximizing the action $\mathcal{S}$ at the exponent of (\ref{FERS})
\begin{equation}
\mathcal{S}=-\frac{n(n-1)}{4}\beta^2\rho^2\sigma^2Sq_0^2-\frac{n}{4}\beta^2\rho^2\sigma^2Sq_D^2+\frac{n}{2}\beta\rho\mu S h^2+\sum_i\ln Z_i
\end{equation}
we get the following saddle point (SP) equations on the introduced parameters 
\begin{equation}
q_0=\overline{\frac{1}{S}\sum_i\langle N_i^a N_i^b\rangle_{\rm AR}}^V
\end{equation}
\begin{equation}
q_D=\overline{\frac{1}{S}\sum_i\langle {N_i^a}^2 \rangle_{\rm AR}}^V
\end{equation}
\begin{equation}
h=\overline{\frac{1}{S}\sum_i\langle N_i^a \rangle_{\rm AR}}^V
\end{equation}
with 
\begin{eqnarray}
  \overline{\langle (N^b)^p (N^c)^r\rangle_{\rm AR}}^V= \phantom{aywrtowyrtowtorwtyorteortyoerytoertyore}\nonumber \\  \phantom{a} \nonumber \\\nonumber
 = \overline{\frac{\int \frac{dz}{\sqrt{2\pi}} \prod_a dN^a \exp\left[-\frac{z^2}{2}-\beta \sum_a H_{\rm RS} (N^a,z) \right](N^b)^p(N^c)^r}{\int \frac{dz}{\sqrt{2\pi}} \prod_a dN^a \exp\left[-\frac{z^2}{2}-\beta \sum_a H_{\rm RS}(N^a,z)\right]}}^V
\end{eqnarray}
\noindent 
where $b,c$ denotes replica indices and $p,r,$ are the powers of the abundance we are considering. 
In the $n\rightarrow 0$ limit the formula above can be expressed in terms of thermal averages $\langle\cdot\rangle_{1{\rm R}}$ over single species and single replica with Hamiltonian $H_{RS}$,
and the disorder average $\overline{\ \cdot\ }$ representing the average over the disorder contained in $V(N)$ and the Gaussian integral over $z$ with mean zero and unit variance
\begin{equation}
  \overline{\langle (N^a)^p (N^b)^r\rangle_{\rm AR}}^V=\overline{\langle N^p\rangle_{1{\rm R}}\langle N^r\rangle_{1{\rm R}}}
\end{equation}
where
\begin{equation}
\langle  \ \cdot \ \rangle_{1{\rm R}}=\frac{\int dN \exp\left[-\beta H_{\rm RS}(N,z)\right] \ \cdot \ }{\int dN \exp\left[-\beta H_{\rm RS}(N,z)\right]} \ .
\end{equation}
Hence we can write 
\begin{equation}
q_0=
\overline{\langle N \rangle^2_{1{\rm R}}}
\end{equation}
\begin{equation}
q_D=
\overline{\langle N^2 \rangle_{1{\rm R}}}
\end{equation}
\begin{equation}
h=
\overline{\langle N \rangle_{1{\rm R}}} \ .
\end{equation}
The average over a single replica coincides with the standard average first over the Boltzmann weight and then over the quenched disorder. It is possible to reduce the latter to the former because the model is mean-field. The resulting equations have a clear interpretation: each species is subjected to
its own potential $V$ and two extra terms due to the overall mean-field interaction with the rest of the system. 
Two non fluctuating terms, one quadratic and the other linear, plus a fluctuating linear term proportional to $z$. The latter can make the minimum of the overall potential zero (extinction) or larger than zero (survival). \\
Note that for fluctuating $K_i$, the LV potential is $V_i(N_i)=-\rho\left(K_i N_i-\frac{N_i^2}{2} \right)$ and the average 
in the 1replica computation above is performed over the Gaussian distributed $K_i$ with mean $\mu_K=1$ and variance $\sigma_K$. 
\subsubsection*{Zero Temperature Limit}
In the zero temperature limit $q_0\rightarrow q_D$ at the same pace as $T$ so it is useful to define the variable $\Delta q=\rho \beta (q_D-q_0)$, where $\rho$ has been inserted just for convenience in writing.
The equations of the three parameters in this limit are hence expressed in terms of $h$, $q_0$, and $\Delta q$.\\
In this limit the thermal averages over the 1replica measure are evaluated by saddle point-method at $N^*$, which is the positive minimum of the Hamiltonian $H_{\rm RS}(N)$ when it exists, or zero. Hence the SP equations become
\[
q_0=\overline{N^*(z)^2} \ \ , \qquad h=\overline{N^*(z)} \ \ , \qquad \Delta q=\rho \overline{
\frac{\theta(N^*(z))}{H_{\rm RS}^{''}(N^*(z))}} \ .
\]
where $\theta(x)$ is the Heaviside function, $\theta(x)=1$ for $x>0$ and zero otherwise.
Note that in the case of last equation on $\Delta q=\overline{\langle N^2\rangle-\langle N\rangle^2}$ we had to Taylor expand $H_{\rm RS}$  for small $T$ separately in the case of extinction $N^*=0$ and survival $N^*>0$.
\\
The LV case is particularly simple since $N^*$ reads
\begin{equation}
N^*(z)=\max\left\{0,\frac{K+z\sigma\sqrt{q_0}-\mu h}{1-\sigma^2\Delta q}\right\} \ .
\end{equation}
Until now $K$ and $z$ were two separate Gaussian variables (with averages $1$ and $0$ and variances $\sigma_K$ and $1$, respectively) over which we are averaging.
Combining together these two variables into $\widetilde{z}$ with $0$ average and variance $1$ we get 
\begin{equation}
N^*=\max\left\{0,\frac{\sqrt{\sigma_K^2+\sigma^2 q_0}}{1-\sigma^2\Delta q}(\widetilde{z}+\Delta)\right\} 
\end{equation}
with value of the random variable $\widetilde{z}$ corresponding to extinction $-\Delta=-\frac{1-\mu h}{\sqrt{q_0\sigma^2+\sigma^2_K}}$.
The expression for $q_0$, $h$, and $\Delta q$ are hence immediately obtained as being
\begin{equation}
q_0=\left(\frac{\sqrt{q_0\sigma^2+\sigma^2_K}}{1-\sigma^2\Delta q}\right)^2w_2(\Delta) \ ,
\end{equation} 
\begin{equation}
h=\frac{\sqrt{q_0\sigma^2+\sigma^2_K}}{1-\sigma^2\Delta q}w_1(\Delta) \ ,
\end{equation} 
and
\begin{equation}
\Delta q=\frac{1}{1-\sigma^2\Delta q}w_0(\Delta) \ ,
\end{equation} 
with 
$$w_i(\Delta)=\int_{-\Delta}^\infty \frac{d\widetilde{z}}{\sqrt{2\pi}}\exp\left[-\frac{\widetilde{z}^2}{2}\right](\widetilde{z}+\Delta)^i\ .$$
These equations coincide with the one obtained by the cavity method in \cite{guyPRE}.
\subsection{One step replica symmetry breaking equation}
In the multiple minima phase the RS solution is unstable, as already checked in \cite{guyPRE}. This implies the existence of multiple equilibria. In order to characterize these equilibria one has to study what kind of RSB solution emerges.\\ 
In the following we consider the 1RSB solution: 
the $n$ replica are divided into $n/m$ groups and $Q_{ab}=q_1$ for $a\ne b$ both in the same group, 
$Q_{ab}=q_0$ for $a,b$ in different groups and $Q_{aa}=q_D$ and $H_a=h$.
Once introduced this {\it ansatz} the computation is similar to the RS one.\\
The free-energy expression is in this case 
\begin{equation}
-\beta n F=\ln \overline{\int dq_0dq_1dq_Ddh
\exp\left[S \mathcal{A}(q_0,q_1,q_D,h)
\right] }^V
\label{FE1RSB}
\end{equation}
with 
\begin{eqnarray}
\mathcal{A}(q_0,q_1,q_D,h)&=&-\rho^2\sigma^2\beta^2\frac{n}{4}[(n-m)q_0^2+(m-1)q_1^2+q_D^2] \nonumber \\
&+&\rho \mu \beta\frac{n}{2}h^2+\frac 1 S \sum_i \ln Z_i
\end{eqnarray}
with an effective Hamiltonian, $H_{eff}(\{N^a\}_i)$, for $Z_i$:
\begin{eqnarray}
&& H_{eff}=  \hspace{-0.15cm}\sum_{a}\hspace{-0.05cm}\left[\rho \mu hN_i^a+V_i(N_i^a)+(T-\lambda)\log N_i^a\right]\hspace{-0.05cm} -\hspace{-0.05cm}\frac{\beta \rho^2 \sigma^2}{2}\hspace{-0.05cm}\times \nonumber\\
&& \left[ q_0\hspace{-0.1cm}\left(\sum_{a}^nN_i^a\right)^2 \hspace{-0.2cm}+\frac{n}{m}(q_1-q_0)\hspace{-0.1cm}\left(\sum_{a}^mN_i^a\right)^2\hspace{-0.2cm}
+ (q_D-q_1)\sum_{a}{N_i^a}^2\right] .\nonumber
\end{eqnarray}
To decouple replicas we exploit standard properties of Gaussian integrals and we get 
\begin{eqnarray}
Z_i=&& \int \frac{dz_i}{\sqrt{2\pi}} \exp\left[-\frac{z_i^2}{2}\right] \prod_{a_B=1}^{n/m}\left(\int \frac{d{z_{B,i}^{a_B}}}{\sqrt{2\pi}}\prod_{a(a_B)} dN_i^a\times \right. \nonumber \\
&& \left. \exp\left[-\frac{{z_{B,i}^{a_B}}^2}{2}-\beta\sum_{a(a_B)}H_{1{\rm RSB}}(N_i^a,z_i,z_{B,i})\right]\right)
\end{eqnarray}
with $a(a_B)\in[(a_B-1)m+1,a_B m]$ and
\begin{eqnarray}
H_{1{\rm RSB}}& ( &N,z,z_B)=-\rho^2 \sigma^2 \beta(q_D-q_1) \frac{N^2}{2}+ (T-\lambda)\log N \nonumber\\
& + &V(N)+\left(\rho \mu h-z_B \rho \sqrt{q_1-q_0} \sigma -z \rho \sqrt q_0 \sigma\right)N
 \ .
\end{eqnarray}
By maximizing the action $\mathcal{A}$ at the exponent of (\ref{FE1RSB})
in the $n\rightarrow 0$ limit we get the following SP equations on the introduced parameters 
\[
h=\overline{\frac{1}{S}\sum_i\langle N_i^{a(a_B)}\rangle_{\rm AR}}^V
\]
\[
q_0=\overline{\frac{1}{S}\sum_i\langle N_i^{a(a_B)}N_i^{a(b_B)}\rangle_{\rm AR}}^V 
\]
\[
q_1=\overline{\frac{1}{S}\sum_i\langle N_i^{a(a_B)}N_i^{b(a_B)}\rangle_{\rm AR}}^V
\]
and
\[
q_D=\overline{\frac{1}{S}\sum_i\langle {N_i^{a(a_B)}}^2\rangle_{\rm AR}}^V
\]
with 


\begin{eqnarray}
  \overline{\langle {N^{b(b_B)}}^p {N^{c(b_B)}}^r{N^{d(d_B)}}^s\rangle_{\rm AR}}^V=\phantom{ywertoertyteorytoewreruitye}  \nonumber \\ \phantom{a}
  \nonumber \\
  =\overline{\frac{
  \int d\mu(z,z_B^{a_B}\hspace{-0.2cm},N^{a(a_B)})
  \exp\hspace{-0.1cm} \left[-\beta \hspace{-0.2cm}\sum\limits_{a(a_B)} \hspace{-0.2cm} H_{1{\rm RSB}}^{a(a_B)}\right] \hspace{-0.1cm}
  {N^{b(b_B)}}^p \hspace{-0.05cm}
  {N^{c(b_B)}}^r \hspace{-0.05cm}
  {N^{d(d_B)}}^s \hspace{-0.1cm}
  }{
  \int d\mu(z,z_B^{a_B}\hspace{-0.2cm},N^{a(a_B)})
  \exp\hspace{-0.1cm} \left[-\beta \hspace{-0.2cm}\sum\limits_{a(a_B)} \hspace{-0.2cm} H_{1{\rm RSB}}^{a(a_B)}\right]
  }}^V \nonumber
\end{eqnarray}
\noindent 
where 
$$d\mu
= \frac{dz\exp\left[-\frac{z^2}{2}\right]}{\sqrt{2\pi}}\hspace{-0.cm}
  \prod\limits_{a_B} \hspace{-0.cm} 
  \frac{dz_B^{a_B}\exp\left[-\frac{{z_B^{a_B}}^2}{2}\right]}{\sqrt{2\pi}}
  \hspace{-0.1cm} 
  \prod\limits_{a(a_B)} 
  \hspace{-0.1cm} dN^{a(a_B)}  
$$ 
and $H_{1{\rm RSB}}^{a(a_B)}=H_{1{\rm RSB}}(N^{a(a_B)},z,{z_B^{a_B}})$. \\
These averages in the $n\rightarrow 0$ limit can be expressed in terms of thermal averages $\langle\cdot\rangle_{1{\rm R}}$ over single species and single replica with Hamiltonian $H_{1{\rm RSB}}(N,z,z_B)$
$$\langle\cdot\rangle_{1{\rm R}}=\frac{\int dN \exp\left[-\beta H_{1{\rm RSB}}(N,z,z_B)\right]\ \cdot\ }{\int dN \exp\left[-\beta H_{1{\rm RSB}}(N,z,z_B)\right]}\ ,$$ averages $\langle\cdot\rangle_{\rm mR}$ over the Gaussian variable $z_B$ with additional weight given by 
$\left(\int dN \exp\left[-\beta H_{1{\rm RSB}}(N,z,z_B)\right]\right)^m$
$$\langle\cdot\rangle_{\rm mR}=\frac{
\int \frac{dz_B}{\sqrt{2\pi}} \exp\left[-\frac{{z_B}^2}{2}\right]\left(\int dN \exp\left[-\beta H_{1{\rm RSB}}(N,z,z_B)\right]\right)^m \hspace{-0.1cm}\cdot}{\int \frac{dz_B}{\sqrt{2\pi}} \exp\left[-\frac{{z_B}^2}{2}\right]\left(\int dN \exp\left[-\beta H_{1{\rm RSB}}(N,z,z_B)\right]\right)^m}$$
 and averages $\overline{\ \cdot\ }^V$ representing the average over the disorder contained in $V(N)$ and the Gaussian integral over $z$ with mean zero and unit variance.
 Using all this we have 
\begin{eqnarray}
  \overline{\langle {N^{b(b_B)}}^p {N^{c(b_B)}}^r{N^{d(d_B)}}^s\rangle_{\rm AR}}^V= \nonumber \\
  \overline{
  \langle
  \langle N^p\rangle_{1{\rm R}}
  \langle N^r\rangle_{1{\rm R}}
  \rangle_{\rm mR}
  \langle
  \langle N^s\rangle_{1{\rm R}}
  \rangle_{\rm mR}
  } \ .
\end{eqnarray}
Hence we can write 
\begin{equation}
q_0=
\overline{\langle\langle N \rangle_{1{\rm R}}\rangle_{\rm mR}^2}
\end{equation}
\begin{equation}
q_1=
\overline{\langle\langle N \rangle_{1{\rm R}}^2\rangle_{\rm mR}}
\end{equation}
\begin{equation}
q_D=
\overline{\langle\langle N^2 \rangle_{1{\rm R}}\rangle_{\rm mR}}
\end{equation}
\begin{equation}
h=
\overline{\langle\langle N \rangle_{1{\rm R}}\rangle_{\rm mR}} \ .
\end{equation}
\subsubsection*{1RSB Zero Temperature Limit}
Also in the case we have to considered rescaled variable in the limit $T\rightarrow 0$: $\rho(q_D-q_1)\beta=\Delta q\sim O(1)$, and the scaling of the replica breaking order parameter $m$ is such that $\beta m$ remains of the order of one.
Hence, in the following we will introduce the notation $\beta m=\widetilde{m}$ and keep $\widetilde{m} \sim O(1)$. \\
In this limit, similarly to the RS case, the SP equations read as follows:
\begin{equation}
q_0=
\overline{\langle N^* \rangle_{\rm mR}^2}
\end{equation}
\begin{equation}
q_1=
\overline{\langle {N^*}^2 \rangle_{\rm mR}}
\end{equation}
\begin{equation}
\Delta q=\rho
\overline{\langle \frac{\theta(N^*)}{H^{''}_{1{\rm RSB}}(N^*)} \rangle_{\rm mR}}
\end{equation}
\begin{equation}
h=
\overline{\langle N^* \rangle_{\rm mR}} \ .
\end{equation}
Previously though the expressions were even simpler because we had to compute averages of the kind 
\[
\int \frac{dz}{\sqrt{2\pi}} \exp[-z^2/2] \frac{\left(\int dN \exp{[-\beta H_{\rm RS}(N,z)]N^l}\right)^k}{\left(\int dN \exp{[-\beta H_{\rm RS}(N,z)]}\right)^{k}}
\]
hence, thanks to the evaluation the integral on $N$ by the saddle point, $\exp[-\beta H_{\rm RS}]$ in the numerator and denominator cancel out.\\ 
In this case instead an additional weight depending on $m$ should be considered next to the Gaussian weight on $z_B$ which comes from $\exp[-\beta mH_{1{\rm RSB}}(N^*(z,z_B),z,z_B)]$ when $N(z,z_B)^*\neq 0$. For the same reason the normalization constant 
is non trivial and must be evaluated.\\
The LV choice of $V$ allows for simple explicit expression of the equations.
In particular as in the LV RS case, we combine the Gaussian variables $z$ and $K$ into $\widetilde{z}$ and for every $z_B$ we get 
\begin{equation}
N^*=\max\left\{0,\frac{\sigma\sqrt{q_1-q_0}}{1-\sigma^2\Delta q}(z_B+\Delta(\widetilde{z}))\right\}
\end{equation}
with the new value of the random variable $z_B$ corresponding to extinction
$$-\Delta(\widetilde{z})=-\frac{\widetilde{z}\sqrt{\sigma_K^2+\sigma^2q_0}+1-\mu h }{\sigma\sqrt{q_1-q_0}} \ .$$
For every given $z_B$ the additional weight involving $H_{1{\rm RSB}}=H_{1{\rm RSB}}(N^*,\widetilde{z},z_B)$ is 
\[
\exp[-\widetilde{m}H_{1{\rm RSB}}]=\exp\left[\frac{\widetilde{m}}{2} \frac{\rho\sigma^2(q_1-q_0)}{1-\sigma^2\Delta q}(z_B+\Delta(\widetilde{z}))^2\right]
\]
when $N^*$ is non null.
Hence the normalization constant is
\begin{eqnarray}
\int\frac{dz_B}{\sqrt{2\pi}} \exp[-z_B^2/2]  \left(\int dN \exp[-\beta H_{1{\rm RSB}}(N,\widetilde{z},z_B)]\right)^m  \nonumber =\\ \nonumber A(\widetilde{z})+D(\widetilde{z})
\end{eqnarray}
with 
\[
A(\widetilde{z})=\hspace{-0.15cm}\int_{-\Delta(\widetilde{z})}^{\infty}\hspace{-0.1cm}\frac{dz_B}{\sqrt{2\pi}} \exp\hspace{-0.1cm}\left[\frac{\widetilde{m}}{2} \frac{\rho\sigma^2(q_1-q_0)}{1-\sigma^2\Delta q}(z_B+\Delta(\widetilde{z}))^2-\frac{z_B^2}{2}\right] 
\]
and 
\[
D(\widetilde{z})=\int_{-\infty}^{-\Delta(\widetilde{z})}\frac{dz_B}{\sqrt{2\pi}} \exp\left[-\frac{z_B^2}{2}\right] \ .
\]
With this in mind and defining 
$$d\mu(z_B;\widetilde{z})=\frac{dz_B}{\sqrt{2\pi}} \exp\left[-\frac{z_B^2}{2}+\frac{\widetilde{m}}{2} \frac{\rho\sigma^2(q_1-q_0)}{1-\sigma^2\Delta q}(z_B+\Delta(\widetilde{z}))^2\right] $$
we can finally write the $1$RSB self consistence equations as follows
\[
h=\int\frac{d\widetilde{z}}{\sqrt{2\pi}} \exp\left[-\frac{\widetilde{z}^2}{2}\right] \frac{B(\widetilde{z})}{A(\widetilde{z})+D(\widetilde{z})}
\]
with 
\[
B(\widetilde{z})=\int_{-\Delta(\widetilde{z})}^{\infty}d\mu(z_B;\widetilde{z})\frac{\sigma\sqrt{q_1-q_0}}{1-\sigma^2\Delta q}(z_B+\Delta(\widetilde{z})) \ ,
\]
\[
q_0=\int\frac{d\widetilde{z}}{\sqrt{2\pi}} \exp[-\widetilde{z}^2/2] \frac{B(\widetilde{z})^2}{(A(\widetilde{z})+D(\widetilde{z}))^2} \ ,
\]
\[
q_1=\int\frac{d\widetilde{z}}{\sqrt{2\pi}} \exp[-\widetilde{z}^2/2] \frac{C(\widetilde{z})}{A(\widetilde{z})+D(\widetilde{z})}
\]
with 
\[
C(\widetilde{z})=\int_{-\Delta(\widetilde{z})}^{\infty}d\mu(z_B;\widetilde{z})
\frac{\sigma^2(q_1-q_0)}{(1-\sigma^2\Delta q)^2}(z_B+\Delta(\widetilde{z}))^2 \ ,
\]
and 
\begin{eqnarray}
\Delta q& = &\rho \beta \overline{\langle\langle N^2\rangle_{1{\rm R}}-\langle N \rangle_{1{\rm R}}^2\rangle_{mR}} \\ \nonumber
& = &\frac{1}{1-\sigma^2\Delta q}\int\hspace{-0.1cm}\frac{d\widetilde{z}}{\sqrt{2\pi}} \exp\hspace{-0.1cm} \left[-\frac{\widetilde{z}^2}{2}\right] \hspace{-0.1cm} \frac{A(\widetilde{z})}{A(\widetilde{z})+D(\widetilde{z})} \ .
\end{eqnarray}
Everywhere we could determine also the $\widetilde{m}$ given by a SP equation, which satisfies the following condition
\begin{eqnarray}
0=&& \widetilde{m}^2(q_1^2-q_0^2)\frac{\rho^2\sigma^2}{4}+\int \frac{dz}{\sqrt{2\pi}}\exp\hspace{-0.1cm} \left[-\frac{z^2}{2}\right] \times \\
&& \left[\log(A(z)+D(z))-{\frac{\rho \widetilde{m}(1-\sigma^2\Delta q)}{2}}\frac{C(z)}{A(z)+D(z)}\right]  \ . \nonumber
\end{eqnarray}
What we do is instead to use $\widetilde{m}$ as a parameter through which we can select minima of the $1$RSB structure at different energy levels.
This allows to compute the number of minima with a given energy, using $\widetilde{m}$ as a parameter conjugated to the energy \cite{monasson1996entropy}. The logarithm of the number of minima divided by $S$ is called configurational entropy. 
It is proportional to the derivative of the free energy with respect to $m$ \cite{monasson1996entropy}.
Note that, by definition of $\widetilde{m}$, the configurational entropy of minima corresponding to the equilibrium in the $1$RSB phase is null.\\
In the $n\rightarrow 0$ and $T\rightarrow 0$ limit the free-energy reads
\begin{eqnarray}
-F& = &\frac{1}{\beta n}\ln \overline{Z^n} \nonumber \\
& = &S\left[-\frac{\rho\sigma^2}{4}[\widetilde{m}\rho(q_1^2-q_0^2)+2 q_1\Delta q]+\frac{\rho\mu}{2}h^2 \right.\\
& + &\left.\frac{1}{\widetilde{m}}\int\frac{dz}{\sqrt{2\pi}} \exp[-z^2/2] \log(A(z)+D(z)) \right] \nonumber
\end{eqnarray}
and the configurational entropy is 
\begin{eqnarray}
S_c& = &-m^2\frac{d}{dm}\left(\frac{1}{n}\ln \overline{Z^n}\right) \nonumber\\
& = &\widetilde{m}^2(q_1^2-q_0^2)\frac{\rho^2\sigma^2}{4}+\int \frac{dz}{\sqrt{2\pi}}\exp[-z^2/2] \times \\
& &\left[\log(A(z)+D(z))-{\frac{\rho \widetilde{m}(1-\sigma^2\Delta q)}{2}}\frac{C(z)}{A(z)+D(z)}\right] \ . \nonumber
\end{eqnarray}
\subsection{Instability of the 1RSB phase and marginality condition for the FRSB phase}
We now study the (in)stability of the 1RSB phase and, more generally, obtain the condition for the stability of RSB phases. \\
To obtain the stability condition we consider a generic $k-RSB$ phase and study fluctuations $\delta Q_{ab}$
only inside the inner blocks of the Parisi matrix. This is the so-called replicon eigenvalue and corresponds physically to fluctuations within a state.  As we shall discuss in the next section, this is directly related to the 
Hessian (stability matrix) around one equilibrium. 
We call $L$ the action
that has to be extremized at the saddle-point and we study its Hessian with respect to the fluctuations described above:
\begin{eqnarray}
\frac{\partial^2 L}{\partial Q_{ a b }\partial Q_{c d }}& = &(\beta \rho \sigma)^2\delta_{\langle a b \rangle,\langle c d \rangle} \\ \nonumber
& - &(\beta \rho \sigma)^4 \left(\overline{\langle N^{a} N^b N^c N^d \rangle} 
-\overline{\langle N^{a} N^b \rangle \langle N^c N^d \rangle} 
\right)
\end{eqnarray}
where the average is done over the effective Hamiltonian. All replica indices belong to the same block of size $m\times m$, one of the inner ones. A single inner block is analogous to a replica $m\times m$ symmetric matrix so the corresponding Hessian matrix can be diagonalized rather easily (see Almeida and Thouless). Within the same block there are three independent matrix elements depending whether some replica index are the same:
\[
P=(\beta \rho \sigma)^2-(\beta \rho \sigma)^4 \left(\overline{\langle (N^{a})^2 (N^b)^2  \rangle }
-\overline{\langle N^{a} N^b \rangle^2}\right)
\]
\[
Q=-(\beta \rho \sigma)^4 \left(\overline{\langle (N^{a})^2 N^b N^c \rangle} 
-\overline{\langle N^{a} N^b \rangle^2}\right)
\]
\[
R=-(\beta \rho \sigma)^4 \left(\overline{\langle N^{a} N^b N^c N^d \rangle} 
-\overline{\langle N^{a} N^b \rangle^2}\right) \ .
\] 
The replicon eigenvalue (see \cite{de1978stability}) is $\lambda=P-2Q+R$ with degeneracy $m(m-1)/3$.
The condition for stability is $\lambda \ge 0$. Marginal stability corresponds to $\lambda =0$. \\
To evaluate the replicon eigenvalue in the $1$RSB phase we need to consider a single block $a_B$ and evaluate
\[
\overline{\langle (N^{a})^2 (N^b)^2  \rangle_{AR} }^V=\overline{\langle \langle N^2\rangle_{1\rm{R}}^2 \rangle_{\rm mR}}
\]
\[
\overline{\langle (N^{a})^2 N^b N^c \rangle_{AR} }^V=\overline{\langle \langle N^2\rangle_{1\rm{R}} \langle N\rangle_{1\rm{R}}^2 \rangle_{\rm mR}}
\]
\[
\overline{\langle N^{a} N^b N^{c} N^d  \rangle_{AR} }^V=\overline{\langle \langle N\rangle_{1\rm{R}}^4 \rangle_{\rm mR}} \ .
\]
Hence the replicon eigenvalue can be expressed in a more transparent way as
\[
\lambda=(\beta\sigma\rho)^2\left[1-(\beta\sigma\rho)^2\overline{\langle\left(\langle N^2 \rangle_{1\rm{R}}-\langle N \rangle_{1\rm{R}}^2\right)^2\rangle_{\rm mR}}\right]
\]
where the second moment of $N^2$ within one single state (or equilibrium) appears.
Using the fluctuation dissipation relation one can rewrite the previous equation in term of single species responses: 
\[
\lambda=(\beta\sigma\rho)^2\left[1-(\sigma\rho)^2\overline{\langle\left(\frac{\partial N }{\partial \xi}\right )^2\rangle} \right]
\]
In the FRSB phase the replicon is exactly zero, this is related to the criticality of the phase \cite{spinglassbeyond}, thus one obtain the equation:
\[
(\sigma\rho)^2\overline{\langle\left(\frac{\partial N }{\partial \xi}\right )^2\rangle}=1
\]
which encodes the marginality condition at finite temperature. 
In the small $T$ limit this computation is analogous to the one performed for $\Delta q$. At the end one gets 
the simpler expression 
\[
(\sigma\rho)^2\overline{\langle\theta(N^*)\left(\frac{1}{H^{''}_{1\rm{RSB}}(N^*)}\right)^2\rangle_{\rm mR}}=1 \ .
\]
or its equivalent expression in terms of single species response
\[
(\sigma\rho)^2\overline{\langle\theta(N^*)\left(\frac{\partial N }{\partial \xi}\right)^2\rangle_{\rm mR}}=1\ .
\]
which leads to eq. (\ref{gmayeq}) of the main text. 
For the usual Lotka-Volterra case in which $V(N)$ is quadratic one gets $H_{1\rm{RSB}}^{''}(N^*)=\rho(1-\sigma^2 \Delta q)$ which does not depend on $N^*$. Thus the replicon eigenvalue and consequently the marginality condition is particularly simple:
\[
\overline{\langle\theta(N^*)\rangle_{\rm mR}}\frac{\sigma^2}{(1-\sigma^2 \Delta q)^2}=1
\]  
As spotted earlier, the expression of $\Delta q$ in terms of correlation function is very similar.
In the LV case the similarity becomes even closer:
\begin{eqnarray}
\Delta q& = &\rho\beta\overline{\langle\langle N^2 \rangle_{1\rm{R}}-\langle N \rangle_{1\rm{R}}^2\rangle_{\rm mR}} \nonumber \\
& = &
\rho\overline{\langle\theta(N^*)\frac{1}{H_{1\rm{RSB}}^{''}(N^*)}\rangle_{\rm mR}} \\ \nonumber
& = &
\overline{\langle\theta(N^*)\rangle_{\rm mR}}\frac{1}{(1-\sigma^2 \Delta q)} \ .
\end{eqnarray}
Using together the equation on $\Delta q$ and the marginality condition one finds two simpler appealing expressions:
\[
\phi\sigma^2=\frac 1 4 \qquad \sigma^2 \Delta q=\frac 1 2 
\] 
where $\phi= \overline{\langle\theta(N^*)\rangle}$. 
The first equation is the the limit of stability given by the May Bound: the fraction of surviving species in any equilibrium should be such that the Wigner semi-circle touches zero. The second is a general result valid in the marginal phase. These analytical predictions have been tested in Fig.2 and Fig.3 of the main text.  
\subsection{Phase diagram and numerical solution of the mean-field equations}
\begin{figure}[h!]
\centering
\includegraphics[scale=0.5]{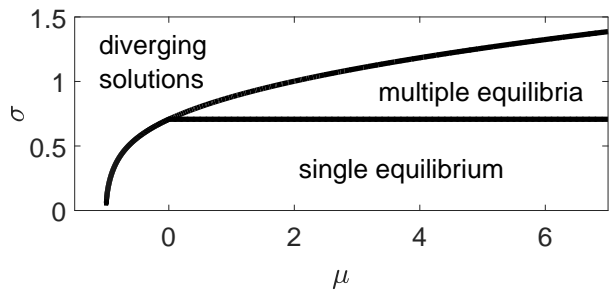}
\caption{Phase Diagram obtained analytically from the mean-field equations.}
\label{phasediag}
\end{figure}

\begin{figure}[h!]
\centering
\includegraphics[scale=0.6]{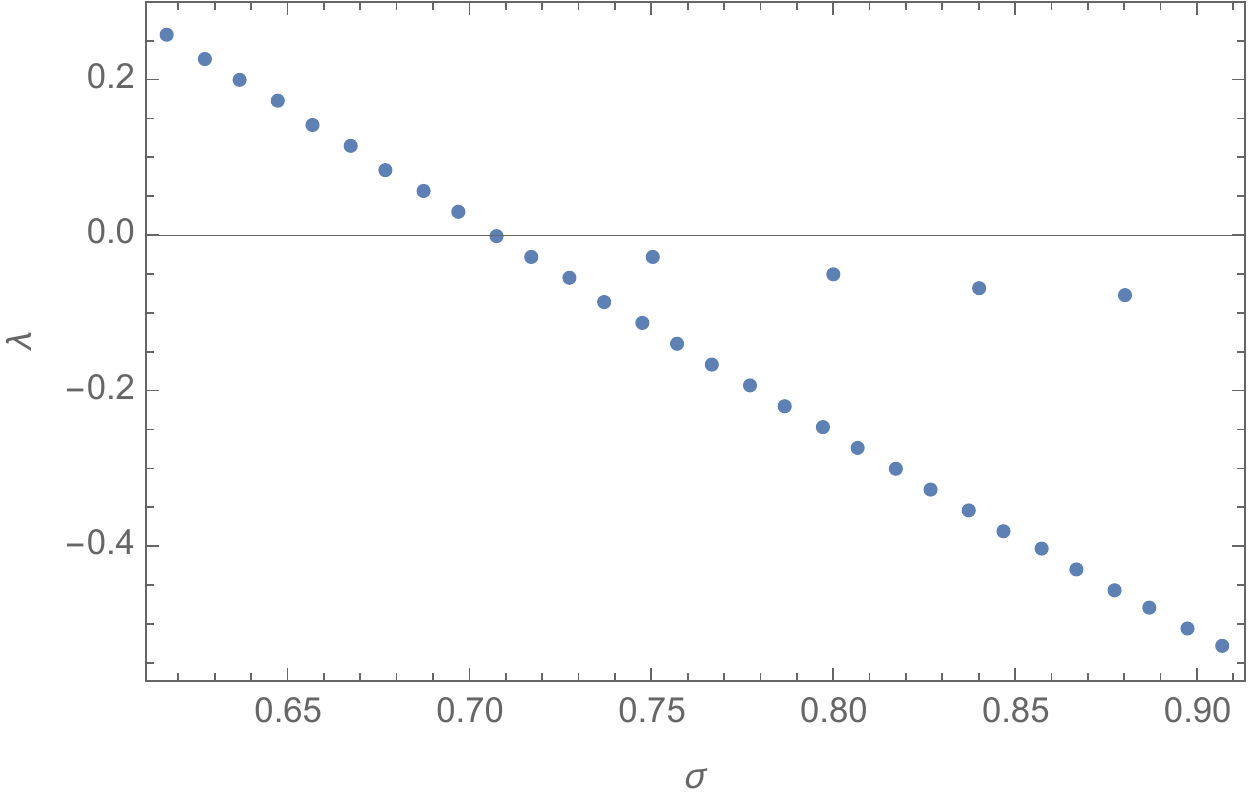}
\caption{Replicon eigenvalue plotted for the RS and the 1RSB phases for $\mu=2$
as a function of $\sigma$. The 1RSB eigenvalue corresponds to the topmost points for $\sigma>\sigma_c=1/\sqrt{2}\simeq 0.707$.}
\label{repliconRS}
\end{figure}

\begin{figure}[h!]
\centering
\includegraphics[scale=0.6]{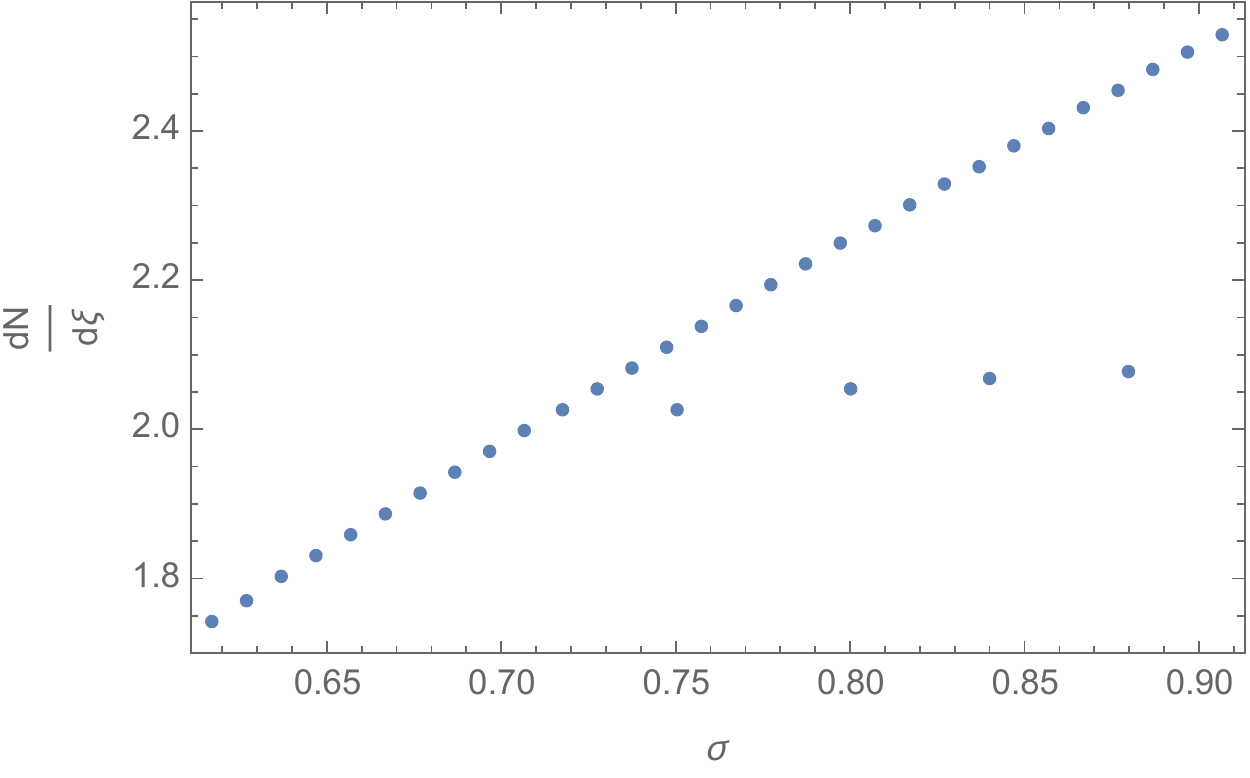}
\caption{Single species response as a function of $\sigma$ for the standard LV model with $r_i=K_i=1$ for $\mu=2$ from the RS and 1RSB solution. The 1RSB result corresponds to the bottom points for $\sigma>\sigma_c=1/\sqrt{2}\simeq 0.707$. The correct FRSB result is $\frac{\partial N_i^*}{\partial \xi_i^*}=2$ for $\sigma>\sigma_c$.}
\label{ssr}
\end{figure}

\begin{figure}[h!]
\centering
\includegraphics[scale=0.6]{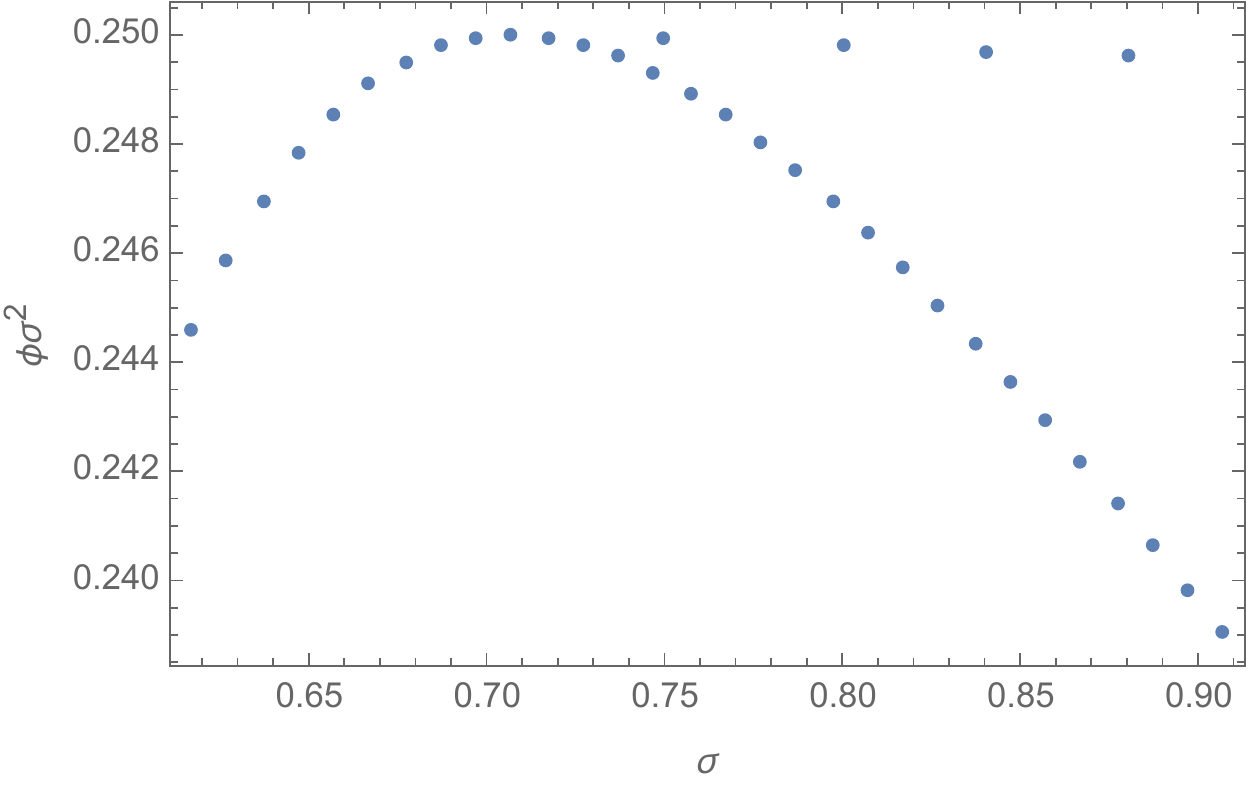}
\caption{$\phi \sigma^2$ as a function of $\sigma$ for the standard LV model with $r_i=K_i=1$ for $\mu=2$ from the RS and 1RSB solution. For $\sigma>\sigma_c$ both phases are unstable but the 1RSB result is very close to the correct 
one corresponding to $\phi \sigma^2=1/4$.}
\label{may-limit}
\end{figure}

The replica symmetric phase was already studied in \cite{guyPRE}. Our results agree with the previous one. In particular we find three phases, see Fig. \ref{phasediag}. By increasing $\sigma$ for $\mu>0$ the single equilibrium phase becomes 
unstable toward the multiple equilibria (spin-glass) phase when its replicon eigenvalue vanishes. 
The instability toward the unbounded growth phase is signalled by a concomitant divergence of $\overline {\langle N\rangle}$ and  $\overline{ \langle N^2\rangle-\langle N\rangle^2}$. 
Note that the transition line to the unbounded growth phase was determined within the RS ansatz so it is only an approximation for $\mu>0$. We have checked by numerical simulations that it is actually a good approximation. \\
Crossing the transition toward the multiple equilibria phase one finds that the RS phase becomes unstable and one
has to break replica symmetry.  We have found that also the 1RSB solution is unstable even though much less than the RS one, see Fig. \ref{repliconRS} where the replicon eigenvalue is plotted for the RS and the 1RSB phases for the standard LV model with $r_i=K_i=1$ for $\mu=2$
as a function of $\sigma$.
We didn't look for 2RSB solutions and directly assumed that the stable phase if the FRSB one as found generically in spin-glass models \cite{spinglassbeyond}. We validated this assumption by comparison with numerical simulations that show marginal stability in the multiple equilibria phase, a property valid only for the FRSB phase. \\
 Note that, although unstable, the 1RSB 
provides a very good approximation as we have checked by comparison with numerical simulations. 
For example, in Fig. \ref{ssr} and \ref{may-limit} we show $\frac{\partial N_i^*}{\partial \xi_i^*}=\frac{1}{1-\sigma^2 \Delta q}$ and $\phi\sigma^2$ for the standard LV model with $r_i=K_i=1$ for $\mu=2$. These two quantities have respectively to stick to the values $2$ and $1/4$ as discussed in the main text and found by numerical simulations. As shown the 1RSB is already a very good approximation 
of the correct results, corresponding to the FRSB phase.  
We have also computed the configurational entropy. Given that the 1RSB is unstable, we cannot determine even approximatively the most numerous equilibria \cite{spinglassbeyond}. The values we found for the configurational entropy as a function of energy within the 1RSB ansatz for the standard LV model with $r_i=K_i=1$ for $\mu=2$ and $\sigma=0.88$ are very small, in the range $10^{-3}-10^{-4}$. It would be interesting (but also quite involved) to obtain 
the correct result within a FRSB computation. Anyhow, it is important to keep in mind that the number of equilibria in realistic situations can be 
modest depending on the value of the configurational entropy and the total number of species. 
\subsection{Random Matrix Analysis}
As explained in the text, the stability of a given equilibrium is governed by the $S^*\times S^*$ stability matrix $M_{ij}^*$ which is defined 
by the equation $(M^*)^{-1}_{ij}=\frac{\partial N_i^*}{\partial \xi_j^*}$ and reads
\begin{equation}\label{SM}
M_{ij}^*=V''(N_i^*)\delta_{ij}+\alpha_{ij}
\end{equation}
In order to study its spectral properties we focus on the resolvent, defined as $G(\lambda)=(S^*)^{-1}\rm{Tr}(\lambda {\bf 1}-M^*)^{-1}$ and add an infinitesimal negative imaginary part to $\lambda$. Following exactly the same procedure developed for mean-field spin-glasses in \cite{bray1979evidence}, one can construct a perturbative expansion in $\alpha_{ij}$, sum the leading contributions for $S\rightarrow \infty$ and obtain the equation:
\begin{equation}\label{gii}
\left(\frac{1}{\lambda {\bf 1}-M^*}\right)_{ii}=\frac{1}{\lambda-V''(N_i^*)-\sigma^2 \phi G(\lambda)}
\end{equation}
valid only for indices corresponding to surviving species. 
By summing over the surviving species one finds 
\begin{equation}\label{RMT2}
G(\lambda)=\langle\frac{1}{\lambda-V''(N_i^*)-\sigma^2 \phi G(\lambda)} \rangle 
\end{equation}
where the average is over the distribution of the $V''(N_i^*)s$. 
This equation allows one to study the 
density of eigenvalues $\rho(\lambda)$ of $M^*$ thanks to the relation $\rm{Im} G(\lambda)=\pi \rho(\lambda)$. 
Since eqs. (\ref{gii},\ref{RMT2}) are also the equations satisfied by the resolvent of a random matrix $\tilde{M}^*$ with independent identically distributed Gaussian off-diagonal entries having the same first and second moment of $\alpha_{ij}$, and independent identically distributed diagonal entries with the same statistics of $V''(N_i^*)$, we conclude that 
$M^*$ and $\tilde{M}^*$ are equivalent as far as the average spectrum is concerned (a relation that we checked 
explicitly by numerics). 
A marginally stable equilibrium is characterized by arbitrary small eigenvalues of its stability matrix, i.e. it is such the left edge of the support of $\rho(\lambda)$ is zero. This implies 
that $\rm{Im} G(\lambda)$ becomes arbitrary small for $\lambda \rightarrow 0$. In consequence, close to $\lambda=0$, we can develop the self-consistent equation on the resolvent as: 
\[
\rm{Im} G(\lambda)=\sigma^2 \phi\langle\left(\frac{1}{\lambda-V''(N_i^*)-\sigma^2 \phi \rm{R}G(\lambda)} \right)^2\rangle  
\rm{Im} G(\lambda)-\]\[-(\sigma^2 \phi)^3 \langle\left(\frac{1}{\lambda-V''(N_i^*)-\sigma^2 \phi \rm{R}G(\lambda)} \right)^4\rangle \left(\rm{Im} G(\lambda)\right)^3+\cdots
\] 
The condition for having a non-zero imaginary part is that when collecting all terms on the RHS the coefficient on the linear term in $\rm{Im} G$ is positive for $\lambda>0$ and vanishes at $\lambda=0$\footnote{Given the type of random matrix we are focusing on, we do not expect any isolated eigenvalue popping out of the spectrum. Therefore, the condition for marginal stability can be obtained from the bulk density of eigenvalues. }. This leads to the equation 
\[
1=\sigma^2 \phi\langle\left(\frac{1}{V''(N_i^*)+\sigma^2 \phi \rm{R}G(\lambda)} \right)^2 \rangle
\]
Using relation (\ref{gii}), and replacing $\left(\frac{1}{V''(N_i^*)+\sigma^2 \phi \rm{R}G(\lambda)} \right)$ by $(M^*)^{-1}_{ii}$ in the identity above, we obtain the equation for marginal stability quoted in the text:
\[{\phi \sigma^2\left(
\frac{1}{S^*}\sum_{i=1}^{S^*}\left((M^*)^{-1}_{ii} \right)^{2}\right)}=1\]

\subsection{Dynamical four-point correlation function $\chi_4(t,t')$}
In the following we derive the analytical results quoted in the main text on $\chi_4(t,t')$ in the limit of small noise.  
First, we define the $S^*\times S^*$ matrix $A$ as
\[
A\equiv\left(  \alpha^{\ast}\right)  ^{-1}%
\]
Let's call also $N_i^*$ the abundance of the surviving species in the limit of zero noise. 
For small noise their abundances have fluctuations of the order $\sqrt{T}$ around the zero-noise value, 
whereas instead abundances of species with $N_i^*=0$ have fluctuations of the order $T$. 
For small noise and at leading order, one can do a quadratic approximation of the Hamiltonian and find for the surviving species:  
\[
\left\langle \delta N_{i}\left(  t\right)  \delta N_{j}\left(  t\right)
\right\rangle =T\newline A_{ij}\
\]
where
\[
\delta N_{i}\left(  t\right)  =N_{i}\left(  t\right)  -N_{i}^{\ast}\ .
\]

The definitions of $C(t,t')$ and $\chi_{4}$ read%
\begin{align*}
C\left(  t,t^{\prime}\right)   &  =\left\langle \frac{1}{S}\sum_{i}\delta
N_{i}\left(  t\right)  \delta N_{i}\left(  t^{\prime}\right)  \right\rangle \\
\chi_{4}\left(  t,t^{\prime}\right)   &  =\frac{S}{C(t,t)^2}\left\langle \left(  \frac{1}%
{S}\sum_{i}\delta N_{i}\left(  t\right)  \delta N_{i}\left(  t^{\prime
}\right)  \right)  ^{2}\right\rangle -S\left[  \frac{C\left(  t,t^{\prime}\right)}{C(t,t)}
\right]  ^{2}%
\end{align*}

The correlation at equal time is related (for small noise) to $\alpha^{\ast}$ via%
\begin{equation}
C\left(  t,t\right)     =\left\langle \frac{1}{S}\sum_{i}\delta N_{i}\left(
t\right)  \delta N_{i}\left(  t^{\prime}\right)  \right\rangle =\frac{T}%
{S}Tr\left[  A\right]  
\end{equation}
Note that the species characterized by zero abundance in the zero noise limit do not contribute at leading order
in $T$ since they would give a contribution $O(T^2)$. 
At long times the correlation function vanishes
\[
C\left(  t\rightarrow\infty,t^{\prime}\right)     =0
\]
This also happens for the correlation between different species:
\[
\left\langle \delta N_{i}\left(  t\right)  \delta N_{j}\left(  t^{\prime
}\right)  \right\rangle =0\ .
\]
so%
\[
\chi_{4}\left(  t,t^{\prime}\right)     =\left(\frac{1}{\frac{T}%
{S}Tr\left[  A\right]  }\right)^2\left\langle S
\left(  \frac{1}{S}\sum_{i}\delta N_{i}\left(  t\right)  \delta N_{i}\left(
t^{\prime}\right)  \right)  ^{2}\right\rangle \overset{t\rightarrow\infty}%
{=}\]
\[
{=} \left(\frac{1}{\frac{T}%
{S}Tr\left[  A\right]  }\right)^2\frac 1 S\sum_{i,j}\left\langle \delta N_{i}\left(  t\right)  \delta
N_{j}\left(  t\right)  \right\rangle \left\langle \delta N_{i}\left(
t^{\prime}\right)  \delta N_{j}\left(  t^{\prime}\right)  \right\rangle \\
\]
\[ =\frac{S}{\left(Tr\left[  A\right]  \right)^2}\sum_{i,j=1}^{S^{\ast}}A_{ij}^{2}=\frac{S}{\left(Tr\left[  A\right]  \right)^2}\sum_{i}\left[  A^{2}\right]  _{ii}=\frac{S Tr\left[  A^{2}\right]}{\left(Tr\left[  A\right]  \right)^2}
\]
while for $t=t^{\prime}$ repeating an analogous computation one finds%
\[
\chi_{4}\left(  t,t\right) =\frac{2S Tr\left[  A^{2}\right]}{\left(Tr\left[  A\right]  \right)^2}  =2\chi_{4}\left(
t\rightarrow\infty,t^{\prime}\right)
\]
The trace can be related to the spectrum via%
\[
\frac{1}{S^*}Tr\left[  A^{n}\right]  =\frac{1}{S^*}Tr\left[  \left(  \alpha^{\ast
}\right)  ^{-n}\right]  =\int d\lambda\frac{\rho\left(  \lambda\right)
}{\lambda^{n}}%
\]
For a semi-circle of radius $a$ centered at $b$,%
\[
\rho\left(  \lambda\right)  =\frac{2}{\pi a}\sqrt{1-\frac{(\lambda-b)^{2}%
}{a^{2}}}\ ,
\]
this gives%
\[
\frac{1}{S}Tr\left[  \left(  \alpha^{\ast}\right)  ^{-2}\right]  =\phi\int
d\lambda\frac{\rho\left(  \lambda\right)  }{\lambda^{2}}=\frac{2\phi}{a^{2}%
}\left(  \frac{b}{\sqrt{b^{2}-a^{2}}}-1\right)  \ .
\]
And for the correlations%
\[
\frac{1}{S}Tr\left[  \left(  \alpha^{\ast}\right)  ^{-1}\right]  =\phi\int
d\lambda\frac{\rho\left(  \lambda\right)  }{\lambda}=\frac{2\phi}{a^{2}%
}\left(  b-\sqrt{b^{2}-a^{2}}\right)
\]

In the standard LV parameterization, the center is at $b=1$ and $a=2\sigma
\sqrt{\phi}$, so one finds
\[
\frac{C\left(  t,t\right)  }{T}   =\frac{1}{S}Tr\left[  A\right]  =\frac
{1}{2\sigma^{2}}\left(  1-\sqrt{1-4\phi\sigma^{2}}\right)
\]
and 
\[
\chi_{4}\left(  t,t\right)  =2\frac{  \frac{1}{\sqrt{1-4\phi\sigma^{2}}%
}-1}{ 1-\sqrt{1-4\phi\sigma^{2}}} \qquad, \qquad \chi_{4}\left(  t\rightarrow\infty,t^{\prime}\right) =\frac{1}{2}\chi_{4}\left(  t,t\right) 
\]
As discussed in the main text and shown in Fig.5, $C(t,t)$ is featureless at the transition while $\chi_4(t,t)$ diverges.



\end{document}